\documentclass[12pt]{iopart}
\usepackage{iopams}       
\usepackage{graphicx}
\usepackage{hyperref}
\usepackage{lipsum}
\usepackage{lettrine}
\usepackage{multirow}
\usepackage{booktabs}
\usepackage{sidecap}
\usepackage{siunitx}
\usepackage{rotating}
\usepackage{orcidlink}
\usepackage{bm}
\expandafter\let\csname equation*\endcsname\relax

\expandafter\let\csname endequation*\endcsname\relax
\usepackage{amsmath}
\usepackage{bbm}  

\def\jqc{Department of Physics, Durham University, Durham DH1 3LE, United Kingdom}
\def\csheffield{Department of Computer Science, University of Sheffield, Sheffield, S1 4DP, United Kingdom}

\begin{document}

\title{Thermometry of simulated Bose--Einstein condensates using {machine learning}}

\author{
Jack Griffiths\orcidlink{0000-0001-7794-6687}\textsuperscript{1,2}, 
Steven A Wrathmall\orcidlink{0000-0003-1770-9721}\textsuperscript{1, *} and 
Simon A Gardiner\orcidlink{0000-0001-5939-4612}\textsuperscript{1}
}

\address{\textsuperscript{1}\jqc}
\address{\textsuperscript{2}\csheffield}

\eads{\mailto{jack.griffiths@sheffield.ac.uk}, \mailto{* Corresponding author s.a.wrathmall@durham.ac.uk}, \mailto{s.a.gardiner@durham.ac.uk}}

\begin{abstract}
Precise determination of thermodynamic parameters in ultracold Bose gases remains challenging due to the destructive nature of conventional measurement techniques and inherent experimental uncertainties. We demonstrate a {machine learning }
approach for rapid, non-destructive estimation of the chemical potential and temperature from {a single image of an} \emph{in situ} imaged density profiles of finite-temperature Bose gases. Our convolutional neural network is trained exclusively on quasi-2D `pancake' condensates in harmonic trap configurations. It achieves parameter extraction within fractions of a second. The model also demonstrates {some} zero-shot generalisation across both trap geometry and thermalisation dynamics, successfully estimating {the temperature (although not the chemical potential)} for toroidally trapped condensates with errors of only a few nanokelvin despite no prior exposure to such geometries during training, and maintaining predictive accuracy during dynamic thermalisation processes after a relatively brief evolution without explicit training on non-equilibrium states. These results suggest that supervised learning can overcome traditional limitations in ultracold atom thermometry, with extension to broader geometric configurations, temperature ranges, and additional parameters potentially enabling comprehensive real-time analysis of quantum gas experiments. Such capabilities could significantly streamline experimental workflows whilst improving measurement precision across a range of quantum fluid systems.


\end{abstract}

\maketitle
 
\section{Introduction}
Since the first experimental realisation of Bose--Einstein condensation in dilute atomic gases \cite{anderson1995observation, bradley1995evidence, davis1995bose} --- one of the most precisely controllable quantum many-body systems --- these systems have become important platforms for quantum simulation \cite{cirac2012goals,bloch2012quantum, schafer2020tools}, precision metrology \cite{cronin2009optics, tino2014atom, bongs2019taking}, and emerging quantum technologies \cite{amico2021roadmap, seaman2007atomtronics}. Central to these applications is the precise thermodynamical characterisation --- particularly values of the temperature and chemical potential --- which, in thermal equilibrium under the grand canonical ensemble, fully determine the system's quantum statistical state and collective behaviour.

Conventionally, the temperature and chemical potential in ultracold atomic experiments are extracted using destructive time-of-flight imaging techniques \cite{Leanhardt2003, Gati2006, Gati2006_2}. In this procedure, the trapping potential is abruptly turned off, allowing the atomic cloud to expand freely; temperature is then deduced from the spatial distribution, under the assumption of a Maxwell--Boltzmann velocity distribution. However, this approach is inherently destructive and suffers from significant shot-to-shot variability, limiting experimental reproducibility and precision. Furthermore, the Maxwell--Boltzmann approximation is fundamentally unsuitable for describing quantum degenerate gases, which exhibit significant quantum statistical correlations and non-trivial interactions.

Non-destructive methods using atomic impurities as temperature probes have recently emerged to address these challenges. Minimally invasive thermometry approaches that use impurity-based polarons have been proposed and shown to be capable of nanokelvin and subnanokelvin precision by analysing impurity fluctuations in momentum and position \cite{Mehboudi2019, Olf2015, Spiegelhalder2009}.

In this paper, we demonstrate a proof-of-principle, non-destructive thermometric technique using {machine learning }
Our method directly estimates temperature and chemical potential from \emph{in situ} density profiles, facilitating repeated, rapid measurements on the same atomic sample without significantly perturbing it. Our approach is inspired by recent advances in machine learning to classify and characterise quantum fluid states \cite{ness, guo, metz, keepfer}. We train our {machine learning} 
model using simulated density distributions generated from the stochastic Gross--Pitaevskii equation (SGPE), a theoretical framework for modelling finite-temperature dynamics and thermal fluctuations in Bose gases. 

We validate our approach across a broad range of temperatures and chemical potentials, and evaluate its estimating capability using unseen density profiles with varied trapping geometries (despite only being trained on harmonically trapped condensates) and during thermalisation (despite only being trained on thermally equilibrated density profiles). Our results demonstrate the robustness and versatility of {machine learning} thermometry, suggesting its potential as a powerful, precise, and experimentally feasible tool for the investigation of ultracold atomic systems.

\section{Description of the physical system}
\subsection{Dynamical treatment of the condensate}
The theoretical treatment of a dilute Bose gas begins with the many-body Hamiltonian
\begin{equation}
 \hat{H}= \int \mathrm{d} \mathbf{x} \hat{\Psi}^{\dagger}(\mathbf{x}) H_\mathrm{sp}(\mathbf{x}) \hat{\Psi}(\mathbf{x})+\frac{1}{2} \iint \mathrm{d} \mathbf{x} \mathrm{d} \mathbf{x}^{\prime} \hat{\Psi}^{\dagger}(\mathbf{x}) \hat{\Psi}^{\dagger}\left(\mathbf{x}^{\prime}\right) U\left(\mathbf{x}-\mathbf{x}^{\prime}\right) \hat{\Psi}\left(\mathbf{x}^{\prime}\right) \hat{\Psi}(\mathbf{x}),
  \label{eq:bec:hamiltonian}
\end{equation}
where
\begin{equation}
  {H}_{\mathrm{sp}}(\mathbf{x})=-\frac{\hbar^2 \nabla^2}{2 m}+V(\mathbf{x})
\end{equation} 
incorporates the single-particle kinetic energy and the external trapping potential $V(\mathbf{x})$, and $U(\mathbf{x-x^\prime})$ represents the two-body interatomic potential, where the factor 1/2 prevents double-counting of particle interactions.

Despite the strong short-range interactions characteristic of alkali atoms, ultracold gases of these species can be treated as weakly interacting systems \cite{burnett}. At very low temperatures, the de Broglie wavelength $\lambda$ greatly exceeds the range of the interatomic potential, permitting the approximation of interactions as elastic, point-like contacts characterised by the $s$-wave scattering length. This leads to the pseudopotential approximation $U(\mathbf{x}-\mathbf{x}^\prime) = g\delta(\mathbf{x}-\mathbf{x}^\prime)$, where the interaction strength $g = {4\pi\hbar^2a_s}/{m}$ is directly proportional to the scattering length $a_s$.

The Heisenberg equation of motion for the Bose field operator $\hat{\Psi}$ is given by
\begin{equation}
  i \hbar \frac{\mathrm{d} \hat{\Psi}(\mathbf{x})}{\mathrm{d} t}=\left[\hat{\Psi}(\mathbf{x}),\hat{H}\right] = H_\mathrm{sp}(\mathbf{x}) \hat{\Psi}(\mathbf{x})+g \hat{\Psi}^{\dagger}(\mathbf{x}) \hat{\Psi}(\mathbf{x}) \hat{\Psi}(\mathbf{x}).
  \label{eq:ch2_hom}
\end{equation} 
We decompose the field operator into condensate and non-condensate contributions \cite{fetter1972}:
\begin{equation}
  \hat{\Psi}(\mathbf{x}) = \hat{\Phi}(\mathbf{x},t) + \hat{\delta}(\mathbf{x},t),
  \label{eq:ch2_field_dec}
\end{equation}
where $\hat{\Phi}(\mathbf{x},t)$ describes the condensate atoms and $\hat{\delta}(\mathbf{x},t)$ represents non-condensate atoms (thermal excitations, quantum fluctuations, or both); unlike the field operator $\hat{\Psi}(\mathbf{x})$, condensate dynamics mean these may both have an explicit time-dependence. In the limit of large condensate occupation, the ensemble average reduces to a classical field: 
$\langle\hat{\Psi}(\mathbf{x})\rangle=\Phi(\mathbf{x},t)$, 
such that 
$\hat{\Psi}(\mathbf{x}) = \Phi(\mathbf{x},t) + \hat{\delta}(\mathbf{x},t)$.

Substituting this decomposition into 
(\ref{eq:ch2_hom}) and taking the expectation value yields a form of nonlinear Schrödinger equation
\begin{equation}
  \begin{aligned} i \hbar \frac{\partial\langle\hat{\Psi}(\mathbf{x})\rangle}{\partial t}  =&i \hbar \frac{\partial \Phi(\mathbf{x},t)}{\partial t} \\ =&H_\mathrm{sp}(\mathbf{x}) \Phi(\mathbf{x},t)+g|\Phi(\mathbf{x},t)|^2 \Phi(\mathbf{x},t)+2 g\left\langle\hat{\delta}^{\dagger}(\mathbf{x},t) \hat{\delta}(\mathbf{x},t)\right\rangle \Phi(\mathbf{x},t) \\  &+g\langle\hat{\delta}(\mathbf{x},t) \hat{\delta}(\mathbf{x},t)\rangle \Phi^*(\mathbf{x},t)+g\left\langle\hat{\delta}^{\dagger}(\mathbf{x},t) \hat{\delta}(\mathbf{x},t) \hat{\delta}(\mathbf{x},t)\right\rangle.\end{aligned}
  \label{eq:ch2_nlse}
\end{equation}
In the limit of negligible fluctuations, this reduces to the Gross--Pitaevskii equation, which provides an accurate description for weakly interacting gases with large atom numbers at temperatures typically below half the critical temperature for Bose--Einstein condensation.

\subsection{Dynamics of Bose gases at finite temperatures}
To investigate Bose gas dynamics across a broader temperature range, including near the phase transition, we require a treatment that incorporates thermal fluctuations beyond the mean-field approximation of 
(\ref{eq:ch2_nlse}).

{We choose to model the growth of the condensate with the stochastic Gross--Pitaevskii equation (SGPE) \cite{duine,stoof1997,stoof2001,stoof1999}, which describes a static coupling of the condensate modes to a thermal bath. The growth is towards a state that corresponds to a thermodynamical equilibrium in the grand canonical ensemble defined by our choice of $\mu$ and $T$, which, by definition, must take real values.} In essence, {the SGPE} is a phenomenologically damped Gross--Pitaevskii equation with additive noise. This approach implements a fluctuation--dissipation theorem that ensures the system relaxes to the correct equilibrium state, without spurious enhancement or depletion of either condensate or thermal components. The SGPE takes the form
\begin{equation}
  i\hbar\frac{\partial \Phi(\mathbf{x},t)}{\partial t} = (1-i\gamma)\left[H_\mathrm{GP}(\mathbf{x}) - \mu \right] \Phi(\mathbf{x},t) + \eta(\mathbf{x},t),
  \label{eq:sgpe}
\end{equation}
where
\begin{equation}
  H_\mathrm{GP}(\mathbf{x}) = -\frac{\hbar^2 \nabla^2}{2m} + V(\mathbf{x}) + g\left|\Phi(\mathbf{x},t)\right|^2
  \label{eq:gpe_hamiltonian}
\end{equation}
is commonly referred to as the Gross--Pitaevskii ``Hamiltonian,'' and the Gaussian noise correlations satisfy
\begin{equation}
  \left<\eta^*(\mathbf{x},t)\,\eta(\mathbf{x}^\prime,t^\prime)\right> = 2\gamma\hbar k_{\mathrm B} T \delta(\mathbf{x}-\mathbf{x}^\prime)\delta(t-t^\prime);
  \label{eq:sgpe_noise}
\end{equation}
note $\eta(\mathbf{x},t)$ is a classical noise term, hence the angle brackets in (\ref{eq:sgpe_noise}) describe a purely statistical averaging, and not an expectation value over a quantum state.

We use a fourth-order Runge--Kutta scheme to solve equation (\ref{eq:sgpe}) on a discretised spatio-temporal grid. There is an implicit projection out of high-momentum modes which are beyond the maximum momentum the spatial grid may represent. Under the numerical scheme as described in equations (\ref{eq:sgpe}--\ref{eq:sgpe_noise}), we do not consider any interactions with the thermal cloud, i.e., we neglect an additive term $2g\tilde{n}(x)$ in equation \ref{eq:gpe_hamiltonian} for the incoherent mean-field contribution (where $\tilde{n}$ is the non-condensate density and the factor of 2 arises from exchange symmetry under a Hartree--Fock theory). Such a term is important for \emph{quantitative\/} agreement with experiments; for further discussion see e.g., \cite{ota, cockburn2}.

Since our focus lies in generating equilibrium thermal states rather than studying formation dynamics, the precise value of the dimensionless coupling parameter $\gamma$ is not critical. This parameter controls how quickly the system equilibrates with its thermal environment---smaller values lead to slower equilibration, while larger values speed it up proportionally. However, $\gamma$ must be chosen carefully: too small and the simulation becomes computationally impractical; too large and topological defects can form, actually slowing convergence. We use a spatially uniform rate $\gamma=0.01$ throughout our simulations, which provides a good balance between computational efficiency and physical accuracy. 

We evolve the system until we reach thermal equilibrium, as monitored by the relative change in condensate number:
\begin{equation}
   \Delta_N = \frac{\int\mathrm{d}\mathbf{x}|\Phi(\mathbf{x},t_n)|^2 - \int\mathrm{d}\mathbf{x}|\Phi(\mathbf{x},t_{n-1})|^2}{\int\mathrm{d}\mathbf{x}|\Phi(\mathbf{x},t_{n-1})|^2}.
   \label{convergence_criteria}
\end{equation}
We assume equilibrium to have been achieved when $\Delta_N<10^{-4}$ for five consecutive time steps.

For machine learning purposes, we partition the resulting atomic density profiles into three datasets: 80\% for training, 10\% for validation during each epoch, and 10\% for final testing. The model is never exposed to the validation or test datasets during training. We use the training set to optimise the model parameters, the validation set to monitor generalisation and prevent overfitting during training, and the test set to provide an unbiased estimate of final performance on unseen data.

\subsection{Atomic species and trap geometry}\label{sec:sec:geometry}
We consider ultracold, single-species atomic Bose gases of rubidium 87 with scattering length $a_s=5.29\,\mathrm{nm}$ \cite{kempen} and atomic mass $1.44\times 10^{-25}\,\mathrm{kg}$. The temperatures are sampled from $T\sim\mathrm{Uniform}(1,200)\,\mathrm{nK}$, encompassing both deeply degenerate and near-critical regimes. Chemical potentials are sampled from $\mu\sim\mathrm{Uniform}(20,80)\,\mathrm{nK}$ rather than from a regular grid to ensure robust model generalisation. Additional parameters are documented in our open-source code \cite{griffiths_mut_model}.

We use highly anisotropic trapping geometries. For harmonic traps, the potential is $V(x, y, z) = m(\omega_x^2 x^2+\omega_y^2 y^2+\omega_z^2 z^2)/2$ with transverse frequencies $\omega_x=\omega_y=2\pi\times 25\,\mathrm{Hz}$ and axial frequency $\omega_z=100\omega_x$. We introduce a slight random anisotropy \footnote{We do this by sampling a random variable $\delta\omega_{x,y} \sim U(-2,2)\mathrm{\,Hz}$ such that the transverse dimensions are $\omega_{x,y}\to\omega_{x,y}+\delta\omega_{x,y}$.} in the transverse dimensions to emulate experimental imperfections. 
We also consider toroidal traps with potential $V(x,y,z) = V_0 \{ 1-\exp( -\sigma^{-2}[\rho(x,y,z)-R]^2 ) \}$, where $V_0$ is the trap depth, $\rho$ is the radial distance from the torus centre, and $\sigma$ and $R$ are the minor and major radii, respectively.

In our model training we use quasi-two-dimensional condensates with an assumed Gaussian profile in the strongly confined $z$-dimension, except in section \ref{model_accuracy}, where we test the model on column-integrated three-dimensional (but still highly anisotropic) condensates. The assumed tight axial confinement is equivalent to requiring $\hbar\omega_z \gg (\mu, k_{\mathrm{B}} T)$ and $\hbar\omega_x, \hbar\omega_y \ll \hbar\omega_z$. The $z$-dimensional contribution is the harmonic oscillator ground state $\phi(z) = (\pi{\hbar}/{m\omega_{z}})^{-1/4}\exp(-m\omega_{z} z^2/2\hbar)$, yielding the complete field $\Phi(x,y,z,t) = \Phi_\mathrm{2D}(x,y,t)\phi(z)$. We can then evolve $\Phi_\mathrm{2D}(x,y,t)$ with a two-dimensional equivalent to (\ref{eq:sgpe}), where $g$ and $\mu$ are replaced by the effective two-dimensional parameters $g_\mathrm{2D} = g \sqrt{m\omega_{z}/2\pi\hbar}$ and $\mu_\mathrm{2D} = \mu - \hbar\omega_z/2$.

\section{Description of the model architecture}\label{sec:architecture}
\subsection{Overview}
\begin{figure*}[ht!]
\includegraphics[width=\textwidth]{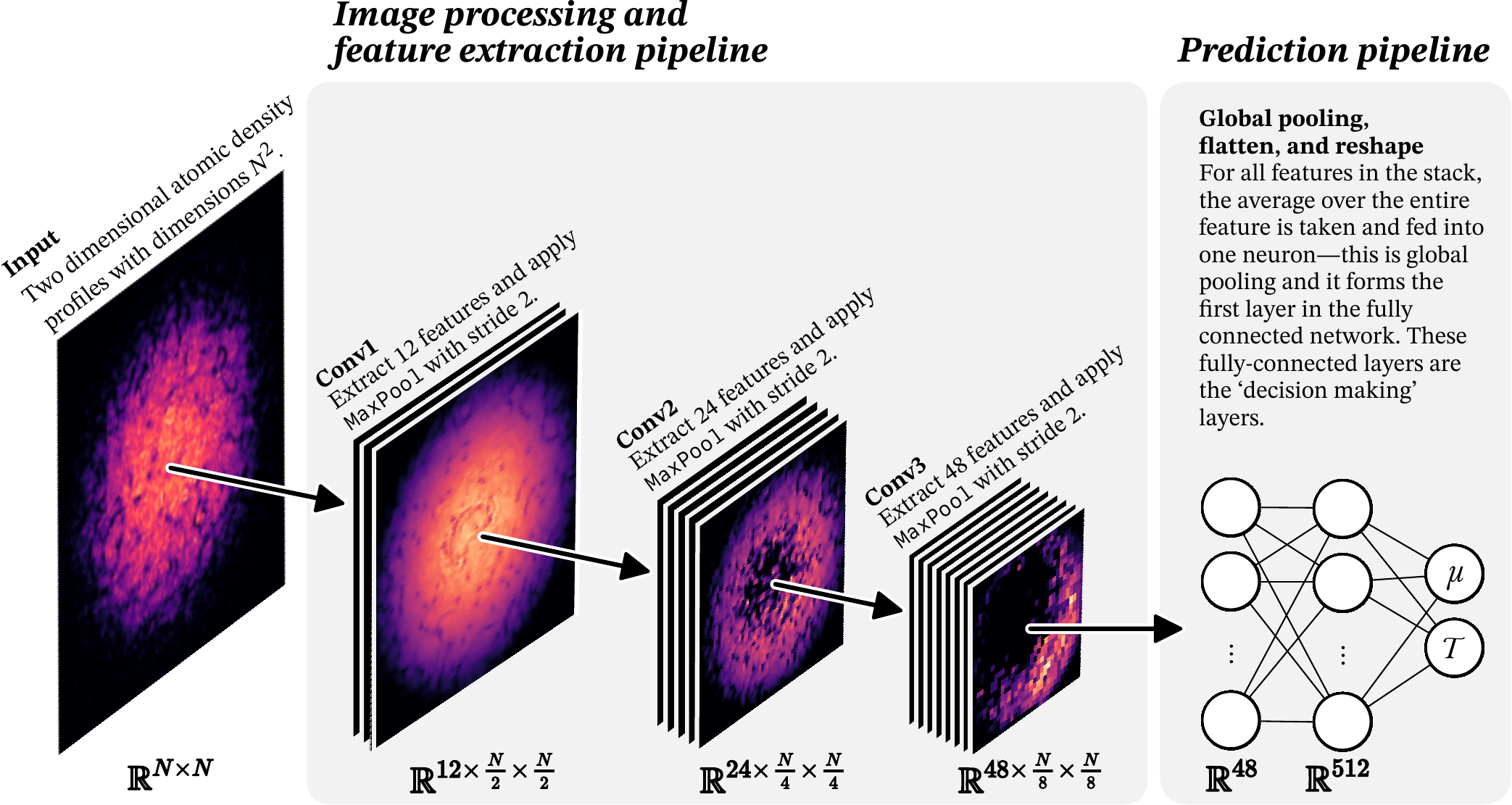}
\caption{Architecture of the convolutional neural network designed to take the density profile of an atomic Bose--Einstein condensate as an input, and from it determine the condensate's chemical potential, $\mu$, and temperature, $T$. The network processes 2D density profiles through three convolutional layers (extracting 12, 24, and 48 features, respectively), each followed by a maximum pooling layer with stride 2 (see \ref{pooling} for details). We apply global pooling to each feature before passing through two fully connected layers (FC1 and FC2), and use ReLU activation functions throughout.}
\label{ch5:architecture}
\end{figure*}

We train our model on a set of $M=3,000$ atomic density profiles, $\rho(\mathbf{x}_\mathrm{2D})$ [where $\mathbf{x}_\mathrm{2D}\equiv(x,y)$], at thermal equilibrium, each of which is a square image of $N$ rows of $N$ pixels, where $N=256$. The network architecture consists of an image processing and feature extraction pipeline followed by a prediction pipeline, as shown in figure~\ref{ch5:architecture}. This architecture effectively processes the rich spatial information in atomic density profiles --- extracting features at various orders of magnitude of the atomic density --- to determine a pair of scalar values which we associate with the chemical potential and temperature.

The image processing and feature extraction pipeline consists of three convolutional layers that increase the number of features (layer 1: 12, layer 2: 24, layer 3: 48) to be extracted in each layer, enabling the model to learn from different aspects of the input data (such as different orders of magnitude of the atomic density or any shape or curvature in the condensate). The reduction of spatial dimensions is not only for computational convenience \footnote{A fully connected neural network for this problem is in principle possible, but at the time of writing would have the same number of weights and biases as the biggest large language models.} --- by reducing the dimensionality of the input, we force the network to learn local, spatially coherent features and reduce the risk of overfitting (by using fewer parameters --- referred to as weights and biases --- in the network) \cite{lecun98}.

The prediction pipeline transforms the spatial features extracted by the convolutional layers into scalar-valued estimations or ``predictions'' of the chemical potential and temperature. By averaging each feature map in Layer 4, we reduce the matrix dimensions and capture global information from each feature. We then expand the model's representational capacity from 48 to 512 neurons in the first fully connected layer (Layer 5 of the overall network) --- this layer facilitates complex, highly non-linear combinations of the pooled features, which is essential for mapping the subtle details of the density profiles to the desired thermodynamic parameters. A large number of neurons in this layer improves the model's predictive and generalisation capabilities --- having fewer neurons in this layer was not as conducive to learning (and sufficiently few meant that the network could not determine the values of $\mu$ and $T$ within any acceptable error) and more neurons did not appreciably improve our model's predictive capabilities. There was no further processing in the output layer (e.g., an activation function); an in-principle consequence is the possibility of predicted negative values for the chemical potential (which can be meaningful, although not in the systems we consider) and the temperature (which is not meaningful). We never observed this to happen, and note that the validity of the SGPE model effectively assumes a finite temperature which is in some sense appreciable.

\subsection{Scaled form of the stochastic Gross--Pitaevskii equation}
When generating the  training data for the model, we consider a scaled system of units in order to avoid computing very small terms (e.g., of the order of $\hbar^2$ expressed in SI units). In particular, we introduce the following scaling factors: {lengths are scaled by the harmonic oscillator lengths $\ell_{x,y,z} = \sqrt{{\hbar}/{m\omega_{x,y,z}}}$}, times by $1/\omega_x$, energies by $\hbar\omega_{x}$, and temperatures by $\hbar\omega_{x}/k_{\mathrm{B}}$. The condensate field is rescaled as $\tilde{\Phi}(\mathbf{x}_\mathrm{2D}, t) = \ell_x \Phi(\mathbf{x}_\mathrm{2D}, t)$, and the position and time variables become $\tilde{\mathbf{x}}_\mathrm{2D} = \mathbf{x}_\mathrm{2D} / \ell_x$ and $\tilde{t} = t / \tau$ respectively, where $\tau = 1/\omega_x$. The Laplacian transforms as $\tilde{\nabla}_\mathrm{2D}^2 = \ell_x^2\nabla^2_\mathrm{2D} $, while the effective two-dimensional chemical potential and interaction strength are rescaled as $\tilde{\mu}_\mathrm{2D} = \mu_\mathrm{2D} / \hbar \omega_x$ and $\tilde{g}_\mathrm{2D} = g_\mathrm{2D} / (\hbar \omega_x \ell_x^2)$. Neglecting the tildes for convenience, the resulting form of the SGPE is:
\begin{equation}
    i\frac{\partial \Phi(\mathbf{x}_\mathrm{2D},t)}{\partial t} = (1 - i\gamma)\left[-\frac{\nabla^2_\mathrm{2D}}{2} + V(\mathbf{x}_\mathrm{2D}) + g_\mathrm{2D}|\Phi(\mathbf{x}_\mathrm{2D},t)|^2 - \mu_\mathrm{2D} \right]\Phi(\mathbf{x}_\mathrm{2D},t) + \eta(\mathbf{x}_\mathrm{2D},t),
\end{equation}
with equivalently scaled Gaussian noise ensemble correlations
\begin{equation}
    \left<\eta^*(\mathbf{x}_\mathrm{2D},t)\,\eta(\mathbf{x}_\mathrm{2D}^\prime,t^\prime)\right> = 2\gamma T \delta(\mathbf{x}_\mathrm{2D}-\mathbf{x}_\mathrm{2D}^\prime)\delta(t-t^\prime).
\end{equation}
It is these forms that we use to generate all atomic density samples in this work.

In our simulations, we consider a dimensionless spatial step size $\Delta x = 0.14262$, which for the $^{87}$Rb system and the $256\times 256$ grid we consider, corresponds to a total grid length of $\SI{80}\um$. We evolve the system by a maximum of 100,000 time steps (although this is typically much smaller depending upon the dynamical time to thermalisation) using a dimensionless time step size $\Delta t = 10^{-3}$. Other simulation parameters (in scaled and unscaled forms) can be found in our source code \cite{griffiths_mut_model}.

In the following sections, we will first describe the prediction pipeline, starting at layer 4 of the overall neural network, followed by a detailed explanation of the image processing and feature extraction pipeline. {We outline the specific mathematical operations used in our convolutional network and contextualise what is often thought of as a `black box,' in a presentation intended to be relevant and accessible for anyone with a physics background.}

\subsection{Prediction pipeline}\label{sec:prediction_pipeline}
\begin{figure}
  \centering
  \includegraphics[width=.6\linewidth]{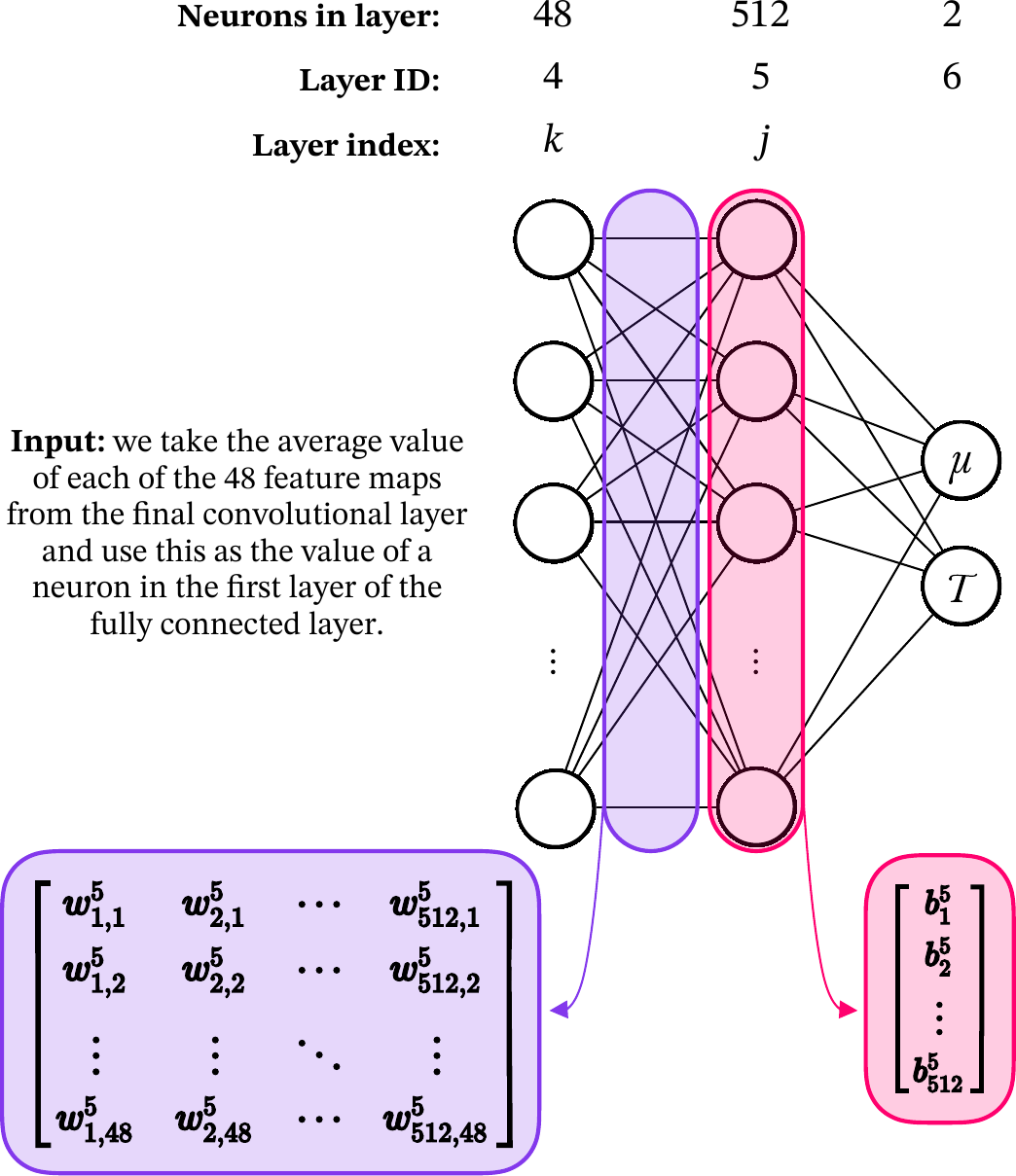}
  \caption{The prediction pipeline architecture, showing in more detail the final part of the network architecture depicted in figure~\ref{ch5:architecture}. Three fully connected, feedforward layers convert the spatial information from the feature extraction pipeline to a pair of scalar values associated with the chemical potential and temperature, highlighting the weights $w_{j,k}^{\ell}$ between layers and the biases $b_{j}^{\ell}$ associated with the neurons in a layer, for $\ell=5$.}
  \label{fig:prediction_pipeline}
\end{figure}

The prediction pipeline is a fully connected, feedforward neural network, consisting of three layers, as shown in figure~\ref{fig:prediction_pipeline}. This is preceded by the feature extraction pipeline, which reduces the dimensionality of the input images, through a sequence of three convolutional layers.
The output of the feature extraction pipeline consists of 48 features $\Xi^3_i$, which are $32\times 32$ square matrices, i.e., $\Xi^3_i \in \mathbb{R}^{N/8 \times N/8}$, where $i \in \{1, \ldots, 48\}$. Taking the average values over the rows and columns of the 48 features $\Xi^3_i$ produces 48 reduced feature representations $y_{i}^{4}$. These are the input values for the first layer of the prediction pipeline, and the fourth layer overall.


In Layer 5, there are 512 neurons with ReLU activation (a necessarily nonlinear function that is commonly used in image recognition scenarios \cite{krizhevsky2012}). The pre-activation values are determined from the $y_{k}^{4}$ through
\begin{equation}
z^5_j = \sum_{k=1}^{48} w^5_{jk} y^4_k + b^5_j, \quad j \in \{1, \ldots, 512\},
\end{equation}
where we define $w^\ell_{jk} \in \mathbb{R}$ as the weight connecting the $k$th neuron in layer $\ell-1$ to the $j$th neuron in layer $\ell$, and $b^\ell_j \in \mathbb{R}$ as the bias for the $j$th neuron in layer $\ell$ (we use the same notation detailed in appendix A of \cite{PhysRevE.111.055302}).
We then apply the activation function to produce
\begin{equation}
y^5_j = \mathrm{ReLU}(z^5_j) = \max(0, z^5_j).
\end{equation}

These values feed into Layer 6 (the final layer), through
\begin{equation}
y^6_j = \sum_{k=1}^{512} w^6_{jk} y^5_k + b^6_j, \quad j \in \{1, 2\},
\label{eq:final_estimate}
\end{equation}
which gives estimations for the chemical potential $\mu = y^6_1$ and temperature $T = y^6_2$, in nanokelvin. 
We initialise the weights and biases using the Kaiming scheme \cite{kaiming}, as $w^5_{jk},\ b^5_j \sim U(-1/\sqrt{512}, 1/\sqrt{512})$ and $w^6_{jk},\ b^6_j \sim U(-1/\sqrt{2},1/\sqrt{2})$ (see appendix C in \cite{PhysRevE.111.055302} for details). {The final layer does not use an activation function, which is motivated by the physics. The chemical potential can be any real value, and the temperature can be any positive, real value, so no activation function is required (one might, in principle, use the $\mathrm{ReLU}=\max(0,x)$ activation for the temperature, but we did not observe negative temperatures being output from our model).\footnote{This contrasts with image classification models, whose final layer typically uses a normalised exponential activation function --- commonly referred to as the `softmax' --- to produce a probability distribution over output classes \cite{bishop2006}.}}

\subsection{Image processing and feature extraction pipeline}\label{sec:x_correl}
\begin{figure}
  \centering
  \includegraphics[width=.6\linewidth]{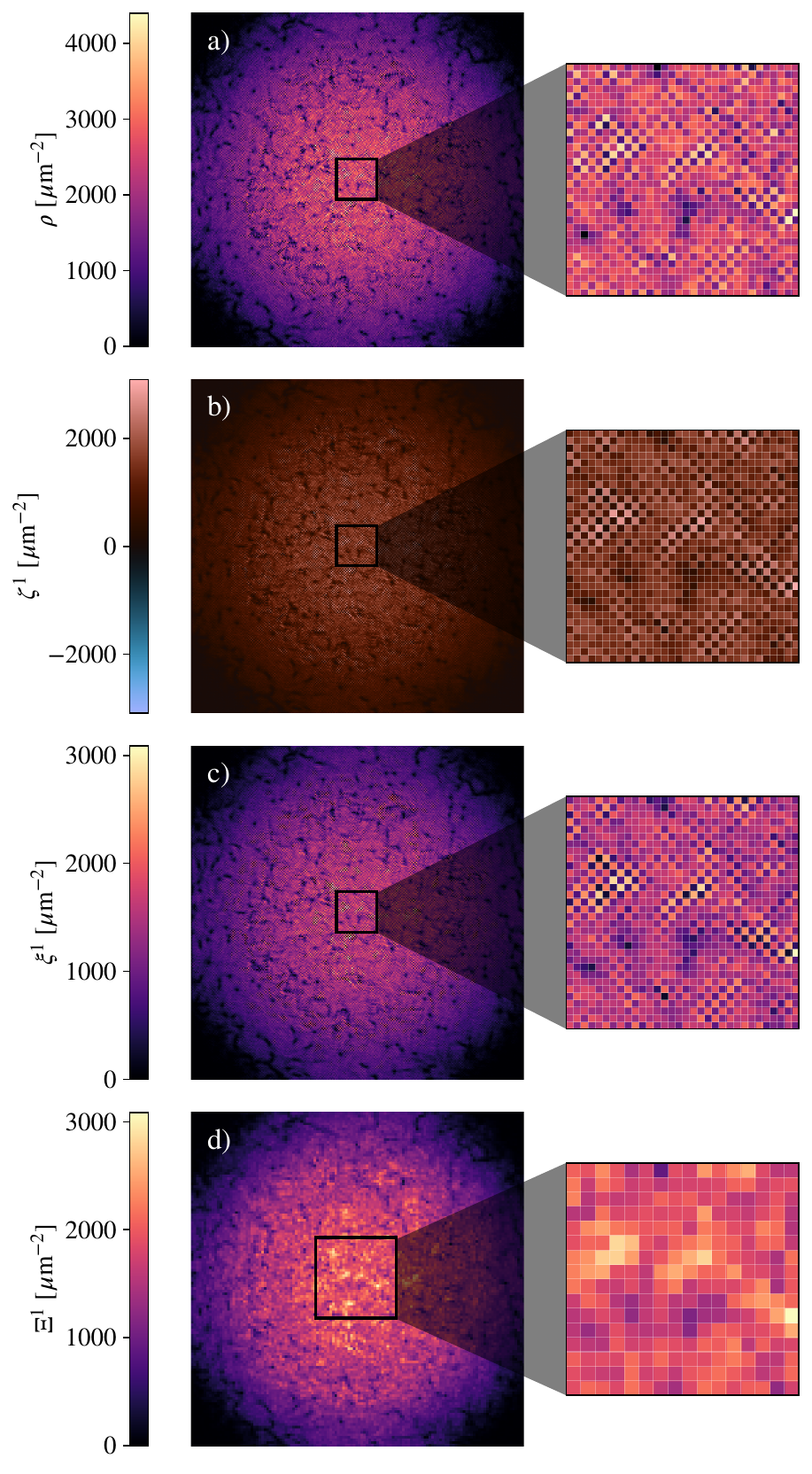}
  \caption{The first convolutional layer \textbf{Conv1}, elaborating in more detail the beginning of the image processing and feature extraction pipeline as depicted in figure~\ref{ch5:architecture}. a) Input of the atomic density, $\rho$. b) Cross-correlation of the atomic density with the weights matrices [see equation~(\ref{eq:layer1_zeta})]. c) Application of the ReLU activation function and a bias [see equation~(\ref{eq:layer1_xi})]. d) Carrying out maximum pooling (resulting in the halving of the image dimensions); this is one of $j=12$ outputs of the first convolutional layer, as per equation~(\ref{eq:layer1_Xi_pool}). To the right we show zoomed-in sections of each step of the first convolutional layer as we construct the first layer feature maps.}
  \label{fig:feature_progression}
\end{figure}

The input atomic density can be thought of as a monochromatic image described by a matrix $\rho \in \mathbb{R}^{N \times N}$. We show an example in figure~\ref{fig:feature_progression} a) --- while we display it using a false colour scale, in terms of information (each pixel has a single value assigned to it, describing the local density) it is functionally monochromatic. The image is processed through three convolutional layers, each followed by ReLU activation and maximum pooling. Figure \ref{fig:feature_progression} demonstrates intermediate steps to calculate a single feature from the first convolutional layer.

\subsection*{Layer 1: From a Single Input to 12 Features}
In the first layer ($\ell=1$), we extract $C_1 = 12$ features from the single input image $\rho$ (such that the number of features in the ``zeroth layer'' is $C_0=1$). The weights for this layer consist of $C_1 = 12$ filters connecting the single input channel ($k=1$) to each output channel $j$, denoted as $\mathsf{W}^{\ell=1}_{j,k=1}(\sigma, \tau)$ for $j=1, \dots, C_1$. Each filter (in this layer and in all subsequent layers) is a square matrix of dimension $3 \times 3$, and we initialise each weight element within these filters, independently, from a uniform distribution $\mathsf{W}^1_{j1}(\sigma, \tau) \sim U(-1/3, 1/3)$ (the range of the probability distribution function depends on the number of features in the previous layer --- see \ref{cnn_weight_init}).

We carry out cross-correlations (see \ref{cross-correlation} for details) between the input image $\rho$ and each filter $\mathsf{W}^1_{j1}$ to compute the intermediate feature maps $\zeta^1_j$:
\begin{equation}
\begin{aligned}
  \zeta^1_j(s_1, t_1) &= (\rho \star \mathsf{W}^1_{j1})(s_1, t_1) \\
  &= \sum_{\sigma=-1}^{1} \sum_{\tau=-1}^{1} \rho(s_1 + \sigma,\ t_1 + \tau)\ \mathsf{W}^1_{j1}(\sigma, \tau), \quad \text{for } j = 1, \dots, 12.
  \label{eq:layer1_zeta}
\end{aligned}
\end{equation}
Here, $s_1, t_1 \in \{1, 2, \dots, 256\}$ (assuming appropriate padding, see \ref{padding}), and we have displayed the cross-correlation of an example atomic density with the final (learnt) weights matrices in figure~\ref{fig:feature_progression} b) (see figure~\ref{fig:ch5:cross_correlation_1} in \ref{appendix:conv_layers} for the outputs of the cross-correlations of the same atomic density with all 12 weights matrices).

We then add a feature-map-specific bias term $b^1_j$ (these are always drawn from the same distribution, appropriate to the layer, as the weights) to each matrix element of the intermediate feature map, and subsequently also apply the ReLU activation function to each element:
\begin{equation}
  \xi^1_j(s_1, t_1) = \mathrm{ReLU}\left(\zeta^1_j(s_1, t_1) + b^1_j \right), \quad \text{for } j = 1, \dots, 12.
  \label{eq:layer1_xi}
\end{equation}
We show the activation function and bias applied to a single output of the cross-correlation in Figure \ref{fig:feature_progression} c) (figure ~\ref{fig:ch5:conv_1} in \ref{appendix:conv_layers} shows the results $\xi_{j}^{1}$ of adding the bias and applying ReLU activation to each of the intermediate feature maps $\zeta_{j}^{1}$).

Finally, we apply maximum pooling with a $2 \times 2$ window and stride 2 (see \ref{pooling} for details) to obtain the first layer's output feature maps $\Xi^1_j$:
\begin{equation}
\Xi^1_j(s_2, t_2) = \max_{\substack{0 \leq \sigma < 2 \\ 0 \leq \tau < 2}} \xi^1_j(2s_2 - 1 + \sigma,\ 2t_2 - 1 + \tau), \quad \text{for } j = 1, \dots, 12.
\label{eq:layer1_Xi_pool}
\end{equation}
This halves the matrix dimensions, so that $s_2, t_2 \in \{1, 2, \dots, 128\}$. These $C_1=12$ feature maps $\Xi^1_j$, each of size $128 \times 128$, become the input to the next layer. We show a single, maximally pooled output in figure~\ref{fig:feature_progression} d)  (figure~\ref{fig:ch5:pool_1} in \ref{appendix:conv_layers} shows the full set of feature maps from the first layer for the particular sample density).

\subsection*{Layer 2: From 12 Features to 24 Features}

In the second layer, we extract $C_2 = 24$ features from the $C_1 = 12$ input feature maps. The weights for this layer consist of $C_2 \times C_1 = 24 \times 12 = 288$ filters, denoted as $\mathsf{W}^2_{jk}(\sigma, \tau)$ for $j=1, \dots, 24$ (output channel index) and $k=1, \dots, 12$ (input channel index). We initialise each weight element within these filters independently, $\mathsf{W}^2_{jk}(\sigma, \tau) \sim  U(-1/\sqrt{108}, 1/\sqrt{108})$.

We compute the intermediate feature maps $\zeta^2_j$ by summing the cross-correlations over all input channels:
\begin{equation}
\begin{aligned}
  \zeta^2_j(s_2, t_2) &= \sum_{k=1}^{12} (\Xi^1_k \star \mathsf{W}^2_{jk})(s_2, t_2) \\
  &= \sum_{k=1}^{12} \sum_{\sigma=-1}^{1} \sum_{\tau=-1}^{1} \Xi^1_k(s_2 + \sigma,\ t_2 + \tau)\ \mathsf{W}^2_{jk}(\sigma, \tau), \quad \text{for } j = 1, \dots, 24.
  \label{eq:layer2_zeta}
\end{aligned}
\end{equation}
See Figure~\ref{fig:ch5:cross_correlation_2} in \ref{appendix:conv_layers} for displays of each output of these cross-correlations from the same sample input. In completely analogous fashion we again add bias terms $b^2_j$, followed by application of the ReLU activation function:
\begin{equation}
\xi^2_j(s_2, t_2) = \mathrm{ReLU}\left(\zeta^2_j(s_2, t_2) + b^2_j\right), \quad \text{for } j = 1, \dots, 24.
\label{eq:layer2_xi}
\end{equation}
Figure~\ref{fig:ch5:conv_2} in
\ref{appendix:conv_layers} equivalently shows the subsequent sample results for all values of $j$. Finally, we apply maximum pooling with stride 2 to obtain the second layer's output feature maps $\Xi^2_j$:
\begin{equation}
\Xi^2_j(s_3, t_3) = \max_{\substack{0 \leq \sigma < 2 \\ 0 \leq \tau < 2}} \xi^2_j(2s_3 - 1 + \sigma,\ 2t_3 - 1 + \tau), \quad \text{for } j = 1, \dots, 24.
\label{eq:layer2_Xi_pool}
\end{equation}
The matrix dimensions again halve, so $s_3, t_3 \in \{1, 2, \dots, 64\}$. These $C_2=24$ feature maps $\Xi^2_j$, each of size $64 \times 64$, become the input to the next layer; Figure~\ref{fig:ch5:pool_2} in  \ref{appendix:conv_layers} displays each of these feature maps resulting from the same sample input density.

\subsection*{Layer 3: From 24 Features to 48 Features}
In the third layer, we extract the final $C_3 = 48$ features from the $C_2 = 24$ input feature maps. 
These final features feed into the subsequent prediction pipeline.
The weights for this layer consist of $C_3 \times C_2 = 48 \times 24 = 1152$ filters, denoted as $\mathsf{W}^3_{jk}(\sigma, \tau)$ for $j=1, \dots, 48$ and $k=1, \dots, 24$. 
We initialise each weight element within these filters independently $\mathsf{W}^3_{jk}(\sigma, \tau) \sim U(-1/\sqrt{216}, 1/\sqrt{216})$.

We carry out cross-correlations, add bias terms, and apply the ReLU activation function and maximum pooling (with stride 2) in exactly the same way as in the second layer. Hence, 
\begin{align}
\begin{split}
  \zeta^3_j(s_3, t_3) &= \sum_{k=1}^{24} (\Xi^2_k \star \mathsf{W}^3_{jk})(s_3, t_3) \\
  &= \sum_{k=1}^{24} \sum_{\sigma=-1}^{1} \sum_{\tau=-1}^{1} \Xi^2_k(s_3 + \sigma,\ t_3 + \tau)\ \mathsf{W}^3_{jk}(\sigma, \tau), \quad \text{for } j = 1, \dots, 48,
  \label{eq:layer3_zeta}
\end{split}
  \\
\xi^3_j(s_3, t_3) & =  \mathrm{ReLU}\left(\zeta^3_j(s_3, t_3) + b^3_j\right), \quad \text{for } j = 1, \dots, 48,
\label{eq:layer3_xi}
 \\ 
\Xi^3_j(s_4, t_4) &= \max_{\substack{0 \leq \sigma < 2 \\ 0 \leq \tau < 2}} \xi^3_j(2s_4 - 1 + \sigma,\ 2t_4 - 1 + \tau), \quad \text{for } j = 1, \dots, 48.
\label{eq:layer3_Xi_pool}
\end{align}
Here, $s_4, t_4 \in \{1, 2, \dots, 32\}$. These $C_3=48$ feature maps $\Xi^3_j$, each of size $32 \times 32$, are the final feature maps from the image processing and feature extraction pipeline that feed into the prediction pipeline, and can be seen in Figure~\ref{fig:ch5:pool_3} in \ref{appendix:conv_layers}, with Figure~\ref{fig:ch5:cross_correlation_3} and \ref{fig:ch5:conv_3} showing the corresponding intermediate steps, for the same sample input density as for each of the previous figures in the Appendix.



\subsection{Computational Considerations and Learning}
\subsubsection{Cost function}
Training a neural network involves optimising its parameters (the weights and biases) to minimise a cost function. For our model, which determines chemical potential and temperature, we consider the square error
\begin{equation}
\mathcal{C} = \frac{1}{2}(\mu^\mathrm{predicted} - \mu^\mathrm{true})^2 + \frac{1}{2}(T^\mathrm{predicted} - T^\mathrm{true})^2.
\label{eq:mut_cost}
\end{equation}
When determining $\mathcal{C}$, we express both the temperature and chemical potential in nanokelvin; with regard to $\mu^\mathrm{predicted}$ and $\mu^\mathrm{true}$, these values are strictly speaking chemical potentials divided by $k_{\mathrm{B}}$, and then expressed in nanokelvin. The values making up $\mathcal{C}$ are thus comparable and of moderate magnitude.

\subsubsection{Batching}\label{sec:batching}
For computational efficiency, we train the network using batches of atomic densities, each with independent associated temperature and chemical potential values, rather than processing them individually. We process input batches of size $\beta$ (typically 16, 32, 64 or 128) as multidimensional arrays\footnote{In the literature, these are referred to as `tensors', although they are not tensors in a physical sense.}, which we form by adding two additional channels, $\beta$ and $C_\ell$, to the matrix at a given layer $\ell$. 
The cost function with batching is
\begin{equation}
\mathcal{C} = \frac{1}{2\beta}\sum_{i=1}^{\beta} \left[(\mu_i^\mathrm{predicted} - \mu_i^\mathrm{true})^2 + (T_i^\mathrm{predicted} - T_i^\mathrm{true})^2\right].
\label{eq:mut_cost_batching}
\end{equation}
Weight matrices and bias terms in each convolutional layer are shared across all items in a batch, such that the same learned filters are applied identically to each input sample. This architectural choice does not reduce the number of parameters relative to processing individual inputs, but it avoids parameter duplication across the batch, thereby preserving model compactness and promoting generalisation. Moreover, while the data are different, because the operations applied to each sample are mathematically identical, the use of batches enables these computations to be vectorised and executed in parallel on GPU hardware. This leads to improved computational efficiency without altering the underlying functional form of the network. In practice, this parallelism is realised by arranging input data as multidimensional arrays with an added batch dimension, facilitating simultaneous convolution, activation, and pooling operations across multiple samples.

\subsubsection{Training}
To minimise the cost function in equation~(\ref{eq:mut_cost_batching}), we use the adaptive moment estimation (Adam) algorithm \cite{kingma} --- a widely used and generally effective adaptive learning rate method suitable for training large networks; see Appendix~D, section 4 of \cite{PhysRevE.111.055302} for detailed derivations --- using the default values for the learning rate, $\eta=0.001$, and decay rates, $(\rho_1, \rho_2)=(0.9, 0.999)$, and backpropagation for gradient calculation. We iteratively use Adam over 100 epochs --- where each epoch processes the entire dataset in randomly selected batches (without replacement) --- although we note \emph{a posteriori} that using fewer epochs is possible since the cost saturates at around epoch 50, as shown in figure \ref{fig:ch5_analysis_validation_A}.
After training, we save the learned weights and biases. Post-training, we can load these optimised parameters into a network with the same architecture to make rapid inferences from new data. We observe inference times for our model of the order of milliseconds in wall-clock time.

As described in section \ref{sec:x_correl}, in each layer the determination of the $\zeta$ matrices is associated with the introduction of the weights, and the determination of the $\xi$ matrices with the introduction of the biases.
Determination of the $\Xi$ matrices is not part of the learning procedure, but processes the images for the next layer by halving the image dimensionality through maximum pooling. 

\begin{figure}
    \centering
    \includegraphics[width=\linewidth]{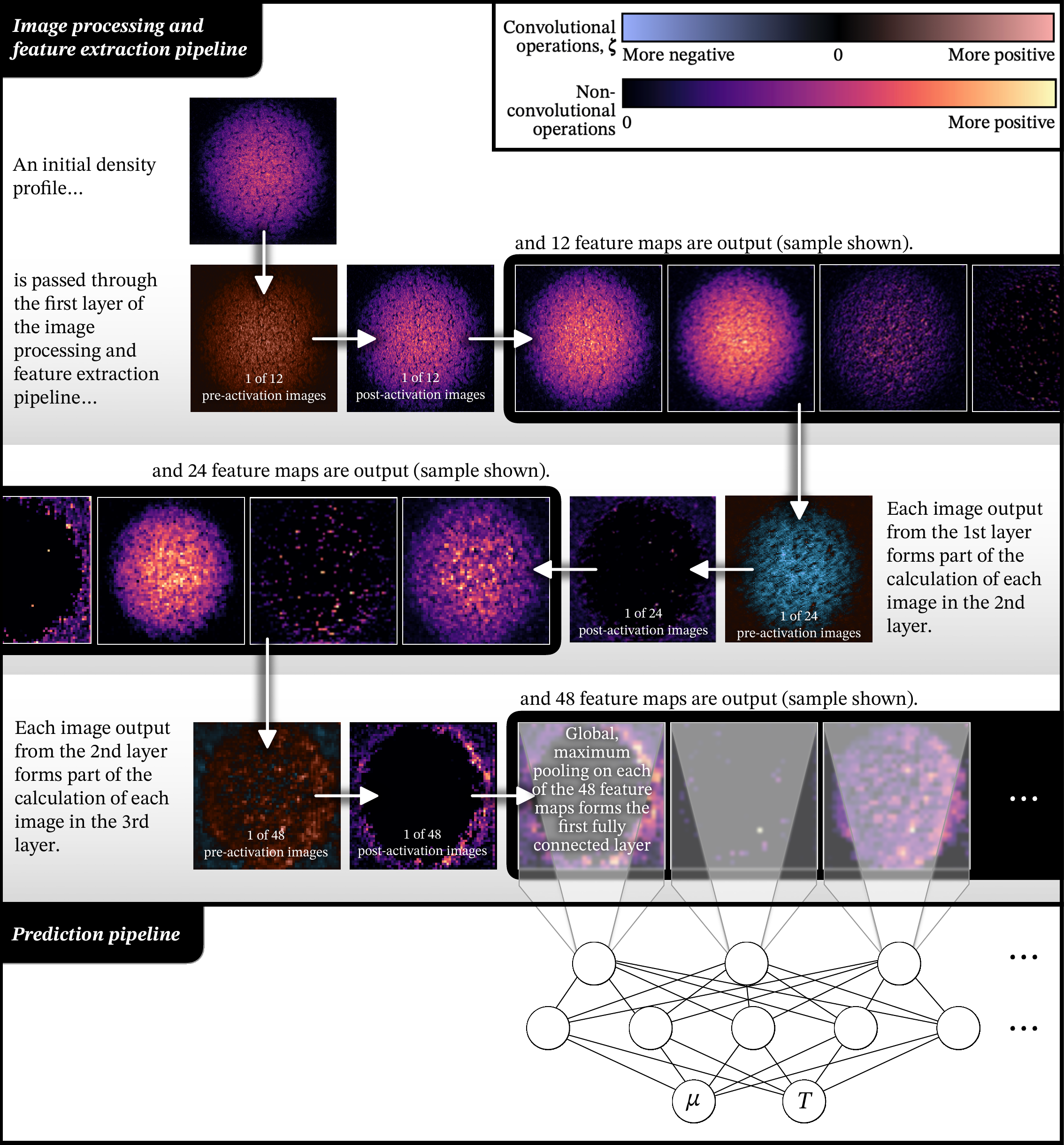}
    \caption{Schematic showing the production of the final 48 feature maps from an initial density profile, via the \textit{image processing and feature extraction pipeline}, which then feed into the \textit{prediction pipeline\/} (see also figure~\ref{fig:prediction_pipeline}) to determine estimated values of the chemical potential $\mu$ and temperature $T$. The initial density profile is the same as that of figure~\ref{fig:feature_progression}(a), the (1 of 12) pre-activation image is the same as in figure~\ref{fig:feature_progression}(b), the following post-activation image is the same as in figure~\ref{fig:feature_progression}(c), and the first of the following sample from 12 feature maps is the same as in figure~\ref{fig:feature_progression}(d). Relative to figure~\ref{ch5:architecture}, this schematic depicts detailed progress, a single channel of each layer at a time, through the image processing and feature extraction pipeline  (see section~\ref{sec:batching} for details). The end-of-layer outputs from all channels (individual ``slices'' in figure~\ref{ch5:architecture}) are inputs to the subsequent layer; the 48 feature maps output by layer 3 are individually globally maximally pooled, as per section~\ref{sec:prediction_pipeline}. The complete set of pre-activation images, post-activation images, and feature maps, for each layer, can be seen in in~\ref{appendix:conv_layers}.}
    \label{fig:feature_progression_full}
\end{figure}

\section{Results and discussion}
\subsection{Interpretation of the feature maps}
In figure \ref{fig:feature_progression_full} we show schematically the steps described mathematically in section~\ref{sec:x_correl} to produce from an initial density profile the feature maps from each of the three convolutional layers in the image processing and feature extraction pipeline. This displays subsets of the 12 feature maps produced in layer 1, the 24 feature maps produced in layer 2, and the 48 feature maps produced in layer 3, where plots, with colour axes, for the complete set of pre-activation images (the $\zeta$ matrices), post activation images (the $\xi$ matrices), and feature maps (the $\Xi$ matrices), for the same run, are displayed in ~\ref{appendix:conv_layers}. 

Increasingly abstract feature maps are learnt as we progress through the three layers. These features may include local variations in the density, patterns, or edges corresponding to the trapping potential, recognisable structures such as vortices, or other structures in the atomic cloud. The features may be local or global, with the convolutional layers trying to identify the fingerprints of the chemical potential and temperature values in the spatial structure of our input samples. 
The first set of feature maps, produced in the first convolutional layer  (a subset of which are shown in figure~\ref{fig:feature_progression_full}), learn high-density features which we can reasonably infer to be associated with the chemical potential, since the chemical potential is fundamentally connected to the average atom number. For the complete set of pre-activation images, post activation images, and feature maps see \ref{appendix:conv_layers}, figures \ref{fig:ch5:cross_correlation_1}, \ref{fig:ch5:conv_1} and \ref{fig:ch5:pool_1}. 
The third and final set of feature maps learn low-density features, which may be associated with thermal fluctuations at this scale. The role of temperature in the atomic density is only introduced in the thermal noise, given by the correlation function in equation~(\ref{eq:sgpe_noise}). Accurate prediction of the temperature using our model is dependent upon being able to effectively capture thermal fluctuations, which the convolutional network is attempting to observe in this final convolutional layer. For the complete set of pre-activation images, post activation images, and feature maps see \ref{appendix:conv_layers}, Figs. \ref{fig:ch5:cross_correlation_3}, \ref{fig:ch5:conv_3}, \ref{fig:ch5:pool_3}.
Between the first and final layers, the second set of feature maps learn a mixture of large-scale features (associated with $\mu$) and small-scale features (associated with $T$); for the complete set of pre-activation images, post activation images, and feature maps see \ref{appendix:conv_layers}, figures \ref{fig:ch5:cross_correlation_2}, \ref{fig:ch5:conv_2}, and \ref{fig:ch5:pool_2}.
This hierarchical feature extraction is essential for accurately predicting both the chemical potential and temperature.

Recognising the distinct roles that the chemical potential and temperature have in our feature maps, we design a series of interrogations to ascertain how our model could be used in a range of experimentally relevant protocols. In particular, we ask: 1) Does the predictive capability of the model depend on whether the density profile being observed is at or very close to thermal equilibrium (which strictly speaking is required for the chemical potential and temperature to be properly defined), e.g., could we predict the chemical potential and temperature during the thermalisation of a condensate? 2) If the model does not have a strong dependence on the equilibrium distribution, could we use the model on toroidally trapped condensates? 3) Can our model predict the chemical potential and temperature of an ensemble average of equilibrium states?
We will address each of these questions in turn, immediately after presenting our choice of model, which we do in the next section.

\subsection{Model accuracy}\label{model_accuracy}

We considered a balance of computational cost (models extracting more features per layer have higher computational cost) and model accuracy (using a metric which we refer to as the accuracy within 5\%, as defined below) to determine which model to use in our subsequent analysis. We considered models with batch sizes of $\beta = 16, 32, 64,$ or $128$ (see section \ref{sec:batching} for details), which extract either 6, 12 and 24 features, 12, 24 and 48 features or 16, 32 and 64 features in each convolutional layer.

\begin{figure}
  \centering
  \includegraphics[width=0.5\linewidth]{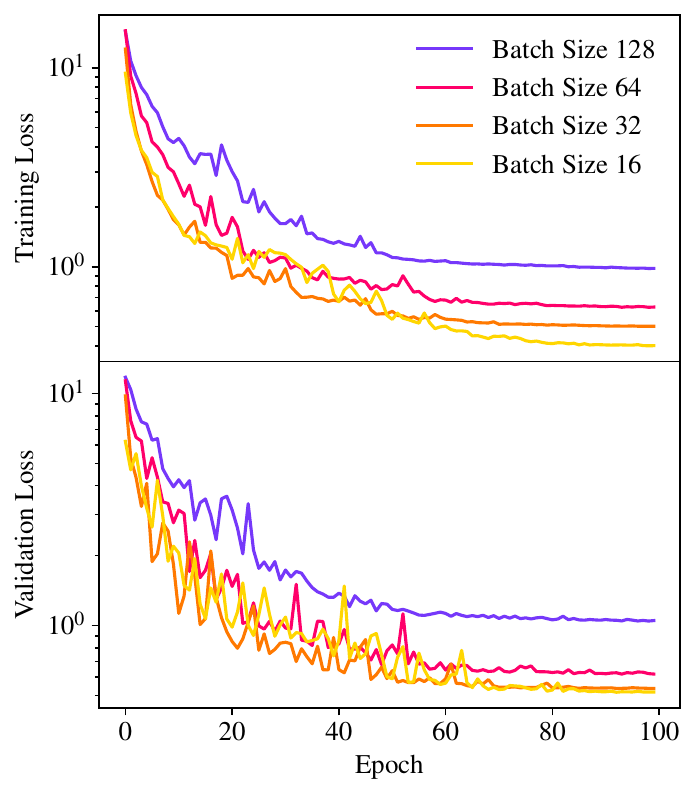}
  \caption{The training and validation cost metrics for batch sizes 16, 32, 64, and 128 from a model extracting 12, 24 and 48 features in the first, second, and third convolutional layers. A smaller batch size results in models with a lower overall training and validation cost.}
  \label{fig:ch5_analysis_validation_A}
\end{figure}

Figure \ref{fig:ch5_analysis_validation_A} shows the training and validation cost metrics for the models that extract 12, 24 and 48 features in the convolutional layers over a range of batch sizes. The absence of divergence between the training and validation cost metrics is often used as a heuristic to assess whether overfitting is occurring; in this case, it provides no such indication. Note that this figure alone is insufficient to ascertain the accuracy of the models. We observe no appreciable divergence in the training or validation cost metrics for all models (with fewer or more features) --- all models approached final costs of the order of $10^{-1}$, despite their \emph{accuracy\/} varying significantly.

\begin{figure}
  \centering
  \includegraphics[width=.6\linewidth]{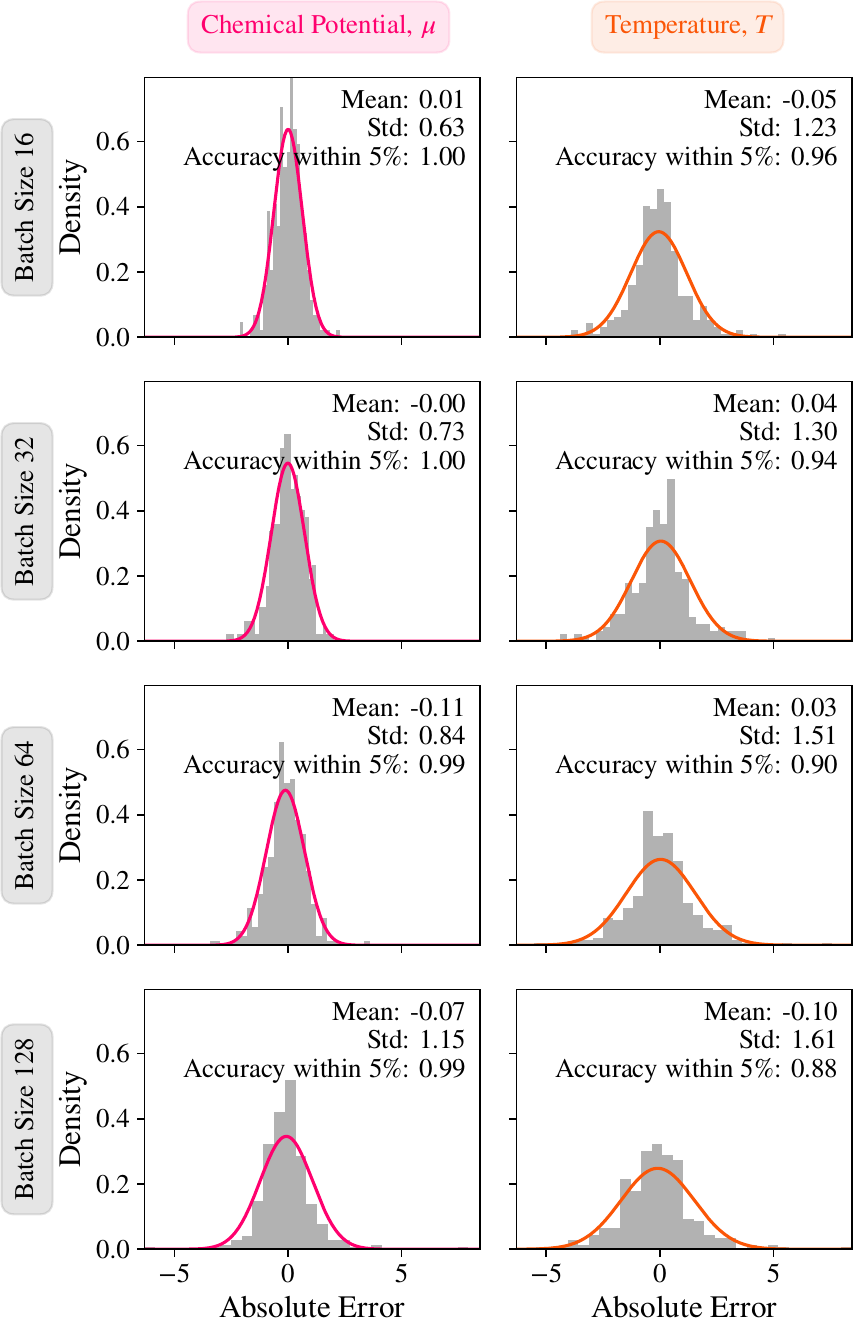}
  \caption[Effect of batch size on classifying $\mu, T$]{The histograms and corresponding normal distributions demonstrate the absolute error in the model's prediction of the chemical potential (column 1) and temperature (column 2) {of the held-out test data set} to the true values input to our SGPE simulations. The plots share a common $x$-axis for easier comparison. These plots correspond to the models which extract 12, 24 and 48 features.}
  \label{fig:ch5_analysis_predictions}
\end{figure}

We define `accuracy within 5\%' as the proportion of samples with predictions within 5\% of their true values --- that is, the values which we set as input parameters to the stochastic evolution. {Using this metric,
figure} \ref{fig:ch5_analysis_predictions} illustrates the predictive accuracy {on the 
test dataset {held back for this purpose}}, 
for models with 12, 24, and 48 convolutional features, across batch sizes $\beta={16, 32, 64}$ and $128$. We see that the standard deviation of absolute errors decreases with decreasing batch size, indicating more accurate predictions of both chemical potential and temperature. Smaller batch sizes generally improve generalisation \cite{keskar} by introducing more stochasticity into the optimisation process, helping the model escape shallow local minima and find solutions that generalise better to unseen data, and this is what we appear to be observing here.

As summarised in Table \ref{tab:accuracy_all_models}, reducing the convolutional feature set to 6, 12, and 24 features has negligible effect on chemical-potential estimation but degrades temperature prediction, whereas the 12, 24, and 48 feature configuration gives the best balance of speed and accuracy. Models with 16, 32, and 64 features produced no measurable improvement over the model with 12, 24, and 48 features ---accuracies unchanged to two decimal places. We have found that the model that extracts 12, 24, and 48 features, with a batch size of 16, offers a good balance between speed and accuracy. All analyses from this point on use this model.
\begin{table}[t]
    \centering
    \caption{{Test dataset accuracy for chemical potential $\mu$ and temperature $T$, 
    defined as the proportion of predictions within $5\%$ of the true values (see \S\,4.2).
    Results shown for three model groups (feature sets) across batch sizes $\beta$.} The models with the best accuracy within 5\% for the chemical potential and temperature are highlighted in bold.}
    \vspace{0.5em}
    \begin{tabular}{lccc}
        \toprule
        Model (features) & Batch size $\beta$ & Accuracy within 5\% ($\mu$) & Accuracy within 5\% ($T$) \\
        \midrule
        \multirow{4}{*}{6, 12, 24} 
            & 16  & 1.00 & 0.92 \\
            & 32  & 0.99 & 0.92    \\
            & 64  & 0.98 & 0.82    \\
            & 128 & 0.98 & 0.73 \\
        \midrule
        \multirow{4}{*}{\textbf{12, 24, 48}}
            & \textbf{16}  & \textbf{1.00} & \textbf{0.96} \\
            & 32  & 1.00 & 0.94 \\
            & 64  & 0.99 & 0.90 \\
            & 128 & 0.99 & 0.88 \\
        \midrule
        \multirow{4}{*}{\textbf{16, 32, 64}}
            & \textbf{16}  & \textbf{1.00} & \textbf{0.96} \\
            & 32  & 1.00 & 0.94 \\
            & 64  & 0.99 & 0.90 \\
            & 128 & 0.99 & 0.88 \\
        \bottomrule
    \end{tabular}
    \label{tab:accuracy_all_models}
    \vspace{0.5em}
\end{table}

We briefly note that column-integrated 3D atomic densities may also be used in our model with similar accuracy to the quasi-2D condensates. Given a three dimensional atomic density profile, $|\Phi(x,y,z)|^2$, the column-integrated density (assuming the imaging is in the $z$-direction) is
\begin{equation}
  n(x,y) = \int^\infty_{-\infty} \mathrm{d}z\, |\Phi(x,y,z)|^2.
\end{equation}
The column-integrated density may then be passed through the machine learning model as previously described with no further post-processing.

\subsection{Prediction capability and robustness}
\begin{figure}[h!]
  \centering
  \includegraphics[width=.7\linewidth]{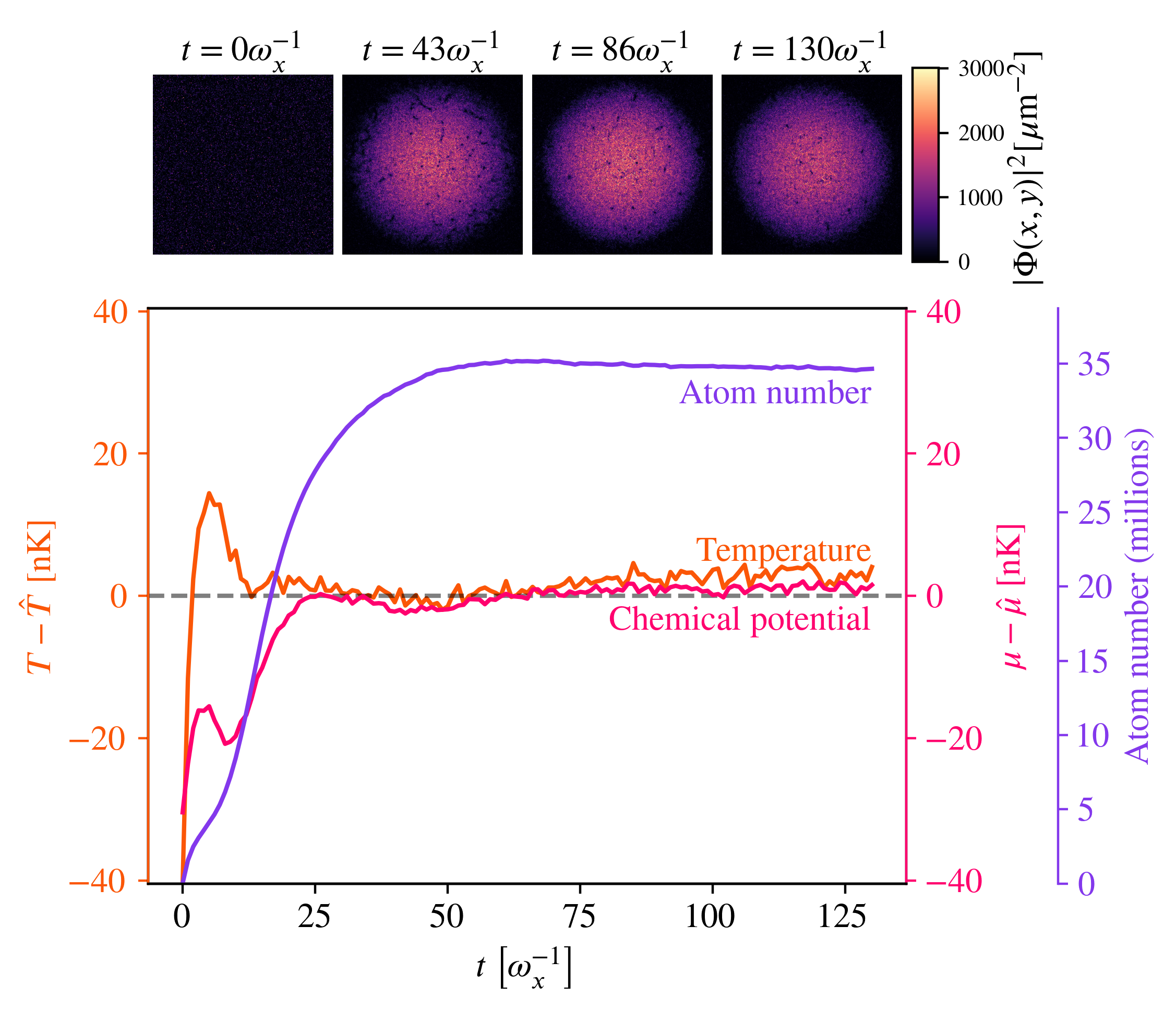}
  \caption[Evolution of the prediction of the chemical potential and temperature during thermalisation for a harmonically trapped condensate]{The evolution of the estimates of the chemical potential and temperature over the thermalisation procedure on dual axes. The pink line is the absolute error in the chemical potential. The orange line is the absolute error in the temperature. The purple line is the atom number. Top: evolution of a harmonically trapped condensate. The temperature of the same is 30 nK. The chemical potential of the sample is also 30 nK. The absolute errors in the temperature and chemical potential plateau after 80 units of rescaled time.}
  \label{ch5:fig_therm}
\end{figure}

\subsubsection{{Estimation capability during thermalisation}}
In our simulations the bath, with which there is formally both heat and particle exchange, is consistently parametrised throughout by a constant chemical potential, $\mu$, and temperature, $T$. We do not consider dynamics following any form of quench, and the machine learning model is trained only on thermalised Bose gases. As described in equation~(\ref{convergence_criteria}), we use convergence of the atom number as a practical criterion for having achieved thermal equilibrium. However we note that estimated values for $\mu$ and $T$ given by our model appear to trend to the final values more quickly, during the thermalisation process {and after a relatively brief transient}, as shown in figure~\ref{ch5:fig_therm}.
The model's ability to estimate parameters during thermalisation, despite training only on equilibrium states, suggests it has learned features that are indicative of the system's eventual thermodynamic state before reaching equilibrium. 


\begin{figure}
  \centering
  \includegraphics[width=.7\linewidth]{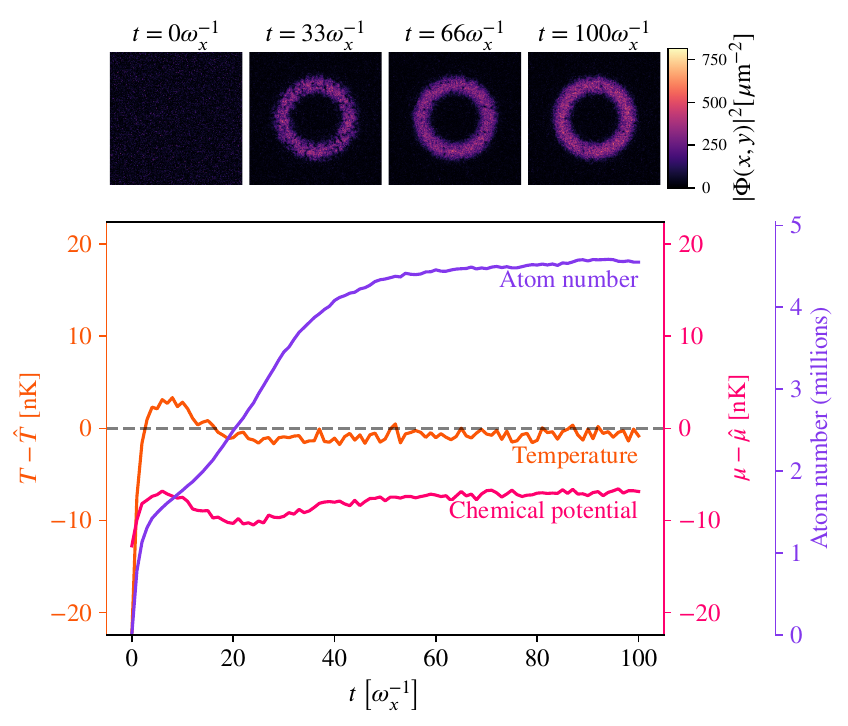}
  \caption[Evolution of the prediction of the chemical potential and temperature during thermalisation for a toroidally trapped condensate]{The evolution of the estimates of the chemical potential and temperature over the thermalisation procedure on dual axes. The pink line is the absolute error in the chemical potential. The orange line is the absolute error in the temperature. The purple line is the atom number. Top: evolution of a harmonically trapped condensate. The temperature of the same is 22 nK. The chemical potential of the sample is also 22 nK. The absolute errors in the temperature and chemical potential plateau after 80 units of rescaled time. The depth of the trap is 60 nK. The minor and major radii are, respectively, $\SI{20}\um$ and $\SI{40}\um$.}
  \label{ch5:fig_therm_torus}
\end{figure}

\subsubsection{{Toroidally trapped condensates}}\label{sec:torus}
Toroidally trapped condensates permit metastable persistent currents (quantised flow of Bose--Einstein condensates in a multiply connected geometry) \cite{ryu3} and provide an appropriate geometry for experimental protocols such as weak links \cite{ramanathan, wright} which are used in several atomtronic devices such as rotational sensors \cite{amico2021roadmap}. As an additional probe of the robustness of our model, which has been trained exclusively with harmonic trapping configurations, we apply it to toroidally trapped condensates. These have depth $V_0=60\mathrm{\,nK}$, minor radius $\sigma=\SI{20}\um$ and major radius $R=\SI{40}\um$; we choose $V_0$ such that $V_0/\mu > 1$.

While, as shown in figure~\ref{ch5:fig_therm_torus}, the estimated $\mu$ and $T$ trend towards stationary values relatively quickly, the temperature is generally closer to the input value than the chemical potential. This difference reflects that $\mu$ is encoded in the bulk density distribution, which differs qualitatively between harmonic and toroidal traps, whereas $T$ is encoded in small-scale fluctuations that are more geometry-invariant.

{Convolutional neural networks exploit weight sharing --- each $3\times 3$ filter
$\mathsf{W}_{jk}^\ell$ is applied identically across all spatial locations, so the same kernel contributes to every receptive field (i.e., each local $3\times 3$ patch of the input). During backpropagation, the gradient of a filter weight is accumulated over all receptive fields, so updating one weight modifies the response of that filter simultaneously at all positions. This enforces a translation-equivariant inductive bias \cite{lecun98}, enabling the network to learn recurring local patterns independently of their absolute position. We interpret that convolutional filters scanning local receptive fields extract temperature-linked signatures that remain invariant under a change from harmonic to toroidal confinement.}

{Two additional factors accentuate the asymmetry: (i) the random transverse anisotropy injected during training (section~\ref{sec:sec:geometry}) encourages robustness of local features to geometric perturbations, and (ii) the maximally global pooling before the fully connected layers discards absolute position, emphasising distributional summaries of fluctuation statistics rather than any geometry-specific structure.}


\begin{figure}
  \centering
  \includegraphics[width=.8\linewidth]{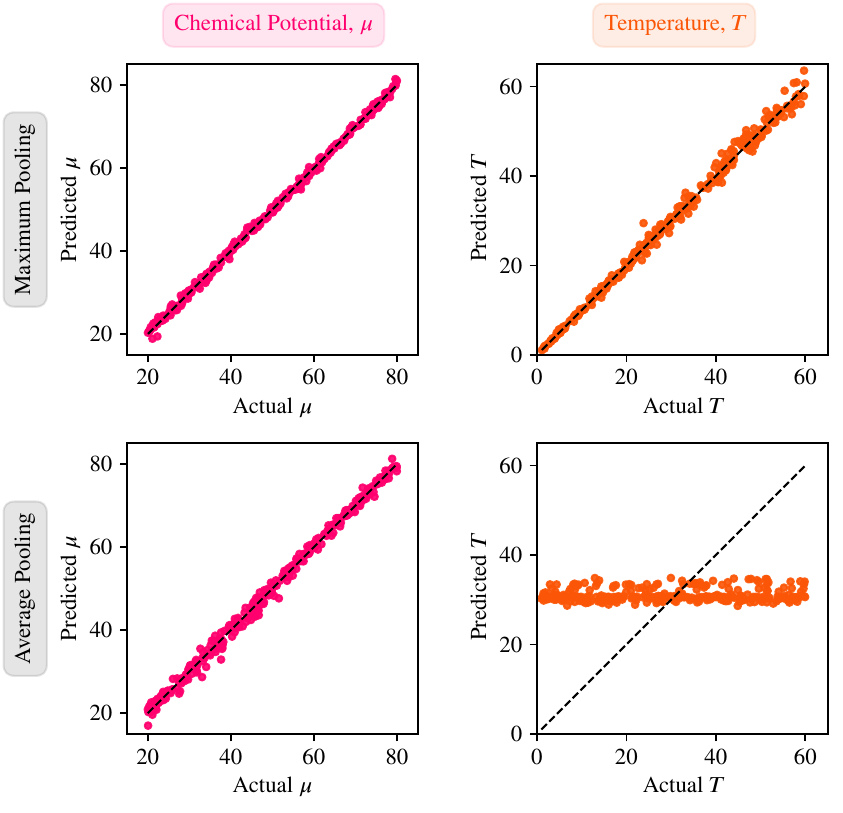}
  \caption{A comparison of the predictive capabilities of a network trained with maximum pooling and a network trained with average pooling. Both maximum pooling and average pooling are appropriate for learning and determining the chemical potential of Bose gases, but only maximum pooling is appropriate for learning and determining their temperature.}
  \label{fig:ch5_max_avg}
\end{figure}
\subsubsection{{Ensemble averages}}
Both average pooling and maximum pooling are commonly used approaches in, e.g., image recognition. As discussed in \ref{pooling}, average pooling gives a general representation of smaller regions in the image whereas maximum pooling emphasises the most prominent features in those regions. It is also possible to think of an equivalency between average pooling and taking an ensemble average over several noise trajectories{, 
where individual trajectories are stochastic;} 
change associated with averaging over these trajectories is smooth and deterministic. Akin to experiments, however, numerical runs obtained from a single noise realisation hold important physical information; one may interpret a single numerical run of the stochastic Gross--Pitaevskii equation as an independent experimental realisation \cite{duine, weiler, cockburn, proukakis}. 

As shown in figure~\ref{fig:ch5_max_avg}, estimation of the chemical potential is quite comparable when using either maximum or average pooling, however when using average pooling information about the temperature is clearly lost. This is consistent with our observation, when considering toroidally trapped condensates, that estimation of the chemical potential is associated with the bulk distribution of the condensate (essentially similar regardless of the pooling type), whereas estimation of the temperature is associated with small-scale local density fluctuations, which will be averaged away by average pooling. 

Similarly, as we have tested, averaging over many trajectories to produce an averaged density washes out the temperature-dependent background fluctuations, and we observe that ensemble averages passed through our machine learning model can only be used to obtain the chemical potential and not the temperature of the sample.

\section{Experimental considerations}
\subsection{{Effective pixel size}}
To establish an effective pixel size in a typical experiment, we consider the imaging system of Wilson, \emph{et al.} \cite{wilson}, which --- like our model --- considers \emph{in situ} imaging of condensates to obtain position distributions. Wilson, \emph{et al.} use a Point Grey Firefly MV CMOS camera with pixels of size $\SI{6.0}\um \times \SI{6.0}\um$ with a magnification of around 19.7, which leads to an effective pixel size of $\SI{6.0}\um/19.7=\SI{0.31}\um$ per pixel. We align our model's {theoretical maximum} resolution to this effective pixel size {based on this camera}. For a quasi-2D `pancake' condensate of diameter $\SI{80}\um$ in a harmonic trap, we need to use a grid size of $256\times 256$ in order to achieve an effective pixel size of about $\SI{0.31}\um$. 

\subsection{{Further considerations}}
{In the work of Wilson, \emph{et al.}, the optical resolution of the microscope (with a numerical aperture of 0.25) is actually diffraction-limited to $\approx1.9$ µm when illuminated with light of wavelength 780 nm. More recent advances in \emph{in situ} imaging of BECs by \cite{gauthier2016} and \cite{kwon2021} report diffraction-limited resolutions of $\SI{0.7}\um$ and $\SI{0.9}\um$, respectively. In practice this means that experimental images are effectively our simulated atomic density profiles convolved with the point-spread function (PSF) of width given by the diffraction-limited resolution of the imaging system. Taking this known complexity into account --- alongside augmentation with the other aforementioned experimental artefacts --- would enhance the model's robustness for real data.} We also note that, while our model has been trained on in situ imaged position distributions, training on expanding time-of-flight velocity distributions is an in-principle straightforward extension, although significantly more demanding to generate if the full dynamics are included.

Whilst our model's resolution is appropriate for experimental data analysis, real BEC images often include additional complexities such as imaging artefacts, focus variations, and dark counts.
{We also draw attention to very recent independent work, developed in parallel by De~Sousa \textit{et al.}~\cite{desousa2025}, for machine learning thermometry of a dilute, laser cooled (non-quantum-degenerate) vapour of potassium 39 atoms, from experimental data.}


\section{Summary}
We have introduced a proof-of-principle machine learning model which can, {with a single image}, estimate the values of the chemical potential and temperature of a Bose-condensed cloud of atoms. We use convolutional neural networks --- the foundation of most image recognition models --- to take an atomic density profile and estimate important thermodynamic parameters.
We have demonstrated that our model for a harmonically trapped condensate can estimate to some degree the chemical potential and temperature for systems that it has not previously been trained on, including during thermalisation, and on toroidally trapped condensates.
This work joins recent applications of machine learning in the quantum fluids literature, such as the identification of topological defects such as vortices and solitons, and the reconstruction of vortex filaments \cite{guo, metz, keepfer}.

The stochastic Gross--Pitaevskii equation was solved in Rust, and is available freely on GitHub \cite{griffiths_sgpe_code}. The machine learning model was written in PyTorch, and is also available freely on GitHub \cite{griffiths_mut_model}.

\section{Data Availability Statement}
The Rust code to reproduce our training data is available at \cite{griffiths_sgpe_code}. The Python code to train our models is available at \cite{griffiths_mut_model}.

\section{Acknowledgements}
We thank K. E. Wilson for insightful discussions about the experimental considerations of this work. We thank the EPSRC, Grant No. EP/T015241/1 and EP/T518001/1, for funding this project.

\appendix

\section{Cross-correlation operation\label{cross-correlation}\label{ch2:cnn}}
In section \ref{sec:x_correl}, we introduced the \textit{cross-correlation\/} function. We define the the cross-correlation function (between $y$ and $w$) as
\begin{equation}
  z(t) = (y\star w)(t) = \sum_{\tau=-\infty}^{+\infty}y(t+\tau)w(\tau).
  \label{eq:x_correl}
\end{equation}
Relative to the more generally known \emph{convolution} operation, the cross-correlation operation is effectively a clockwise rotation of the kernel by 180 degrees. 
We extend equation~(\ref{eq:x_correl}) to matrices by introducing another index, $\sigma$,
\begin{equation}
  \mathsf{Z}(s,t) = (\mathsf{Y}\star \mathsf{W})(s,t)=\sum_{\sigma=-\infty}^{+\infty}\sum_{\tau=-\infty}^{+\infty}\mathsf{Y}(s+\sigma,t+\tau)\cdot \mathsf{W}(\sigma, \tau).
  \label{eq:x_correl_2d} 
\end{equation}
\index{cross-correlation}
The cross-correlation operation may be interpreted as a Frobenius inner product of a sub-matrix of $\mathsf{Y}$ and the kernel, $\mathsf{W}$ \cite{griffiths_thesis}.

For historical and conventional reasons, it is the cross-correlation that is calculated in machine learning libraries such as PyTorch \cite{pytorch-conv2d} and Tensorflow \cite{tensorflow-conv2d} --- ``convolutional'' neural networks are a technical misnomer, but during training, convolutions and cross-correlations should converge on the same result.

Since most of the data we are interested in are two-dimensional matrices, it is equation~(\ref{eq:x_correl_2d}) that we use throughout this paper.

\section{Features, filters, and their initialisation}\label{cnn_weight_init}


In a convolutional layer $\ell$ (where $\ell \in \{1, 2, 3\}$), we aim to extract $C_\ell$ features (output channels) from the $C_{\ell-1}$ features (input channels) provided by the previous layer (layer $\ell-1$). For the first layer ($\ell=1$), the input is the single map $\rho$, so $C_0=1$.

To compute the $j$-th output feature map (where $j \in \{1, \dots, C_\ell\}$), we use a set of weights (filters). Specifically, for each input feature map $k$ (where $k \in \{1, \dots, C_{\ell-1}\}$), there is a 2D filter $\mathsf{W}^\ell_{jk}$ of spatial size $k_W \times k_W$. For $\ell=1$, $k$ can only be 1, so the filters are $\mathsf{W}^1_{j1}$. The collection of $C_{\ell-1}$ filters $\{\mathsf{W}^\ell_{j1}, \dots, \mathsf{W}^\ell_{j,C_{\ell-1}}\}$ is used together (summing their individual cross-correlation results with corresponding input maps) to produce the $j$-th intermediate output map $\zeta^\ell_j$.

We initialise each individual weight element $\mathsf{W}^\ell_{jk}(\sigma, \tau)$ (where $\sigma, \tau$ are spatial indices within the $k_W \times k_W$ filter) independently from the uniform distribution \cite{pytorch-conv2d}:
\begin{equation}
  \mathsf{W}^\ell_{jk}(\sigma, \tau) \sim U\left(-\sqrt{\frac{1}{C_{\ell-1} k_W^2}}, \sqrt{\frac{1}{C_{\ell-1} k_W^2}}\right).
  \label{eq:cnn_weights}
\end{equation}
Here, $C_{\ell-1}$ is the number of input channels (`fan-in' channels) to the layer, and $k_W^2$ is the spatial area of the filter. The term $C_{\ell-1} k_W^2$ represents the total number of inputs that contribute to a single element in the output map before activation (the fan-in). We sample the biases $b_j^\ell$ from the same distribution $U$.

The total number of distinct weight matrices (filters), $N_{\text{filters}}$, in a convolutional neural network with $L$ layers is
\begin{equation}
 N_{\text{filters}} = \sum_{\ell=1}^{L} C_\ell \cdot C_{\ell-1}.
 \label{eq:total_cnn_filters}
\end{equation}

The machine learning problem is to find the appropriate weights $\mathsf{W}^\ell_{jk}$ and biases $b^\ell_j$ which optimise the mapping from the input data to the desired output.

\section{Padding}\label{padding}
To preserve the spatial dimensions of the feature maps during the cross-correlation operation (before pooling), we apply zero padding around the input feature maps of layer $\ell$. For a filter (kernel) of size $k_W \times k_W$, where $k_W$ is odd (e.g., $k_W=3$ as used in the main text), we add $\kappa$ zeros to each side (top, bottom, left, right) of the input maps, where
\begin{equation}
\kappa = \frac{k_W - 1}{2}.
\end{equation}
For $k_W=3$, $\kappa=1$. This `same' padding ensures that the output of the cross-correlation, $\zeta^\ell_j$, has the same matrix dimensions as the input feature maps, $\Xi^{\ell-1}_k$, before pooling is applied.


\section{Pooling}\label{pooling}
\begin{figure}[ht!]
  \centering
  \includegraphics[width=.7\linewidth]{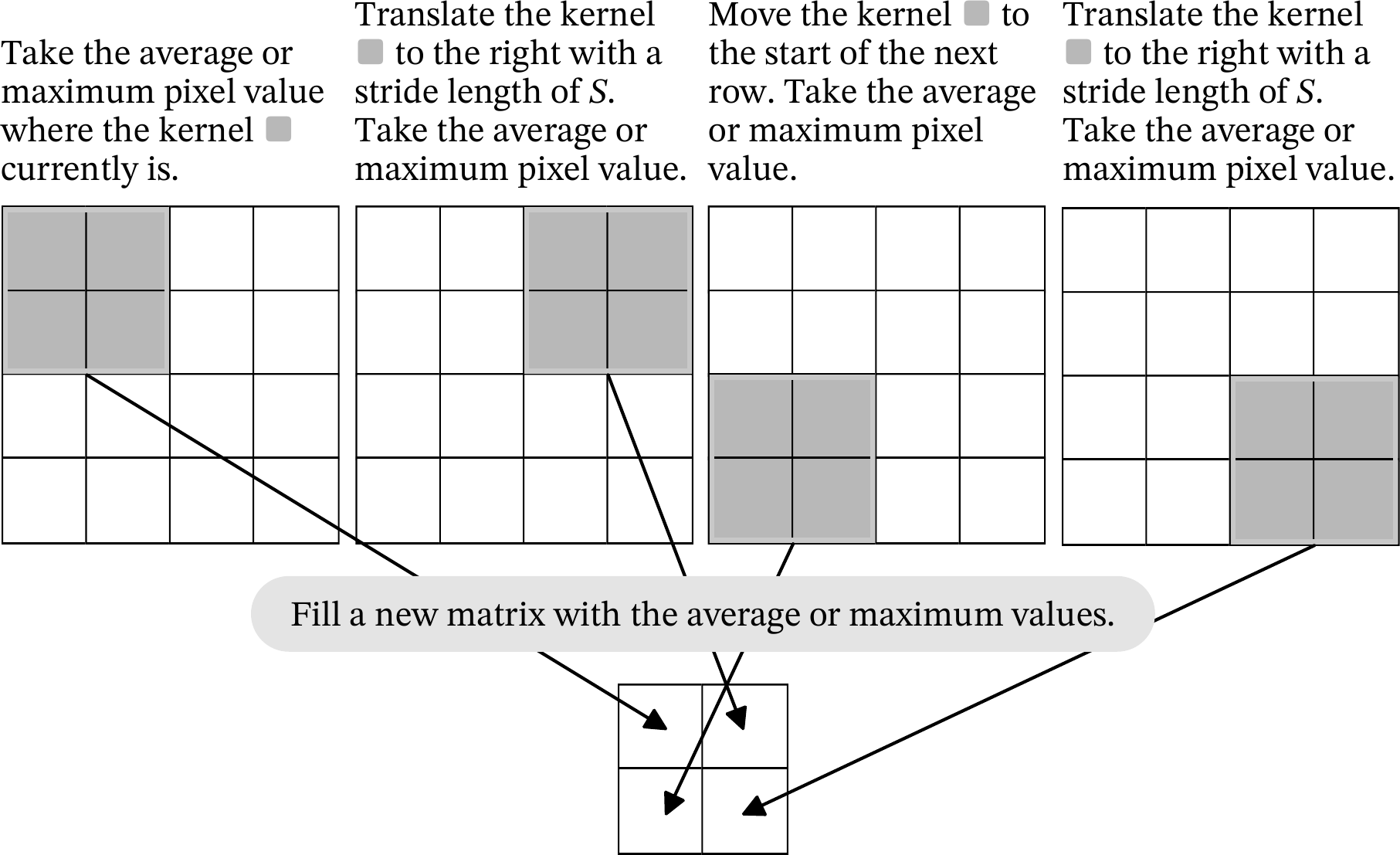}
  \caption[Average or maximum pooling]{Illustration of pooling using a kernel of size $k_P\times k_P=2\times 2$ and stride length $\Sigma=2$ on a $4\times 4$ grid. The output is a square matrix with reduced width and height $\lfloor (N_{in}-k_P)/\Sigma \rfloor + 1 = \lfloor (4-2)/2 \rfloor + 1 = 2$. In the main text, we use maximum pooling with $k_P=2$ and $\Sigma=2$. The convolutional filters have size $k_W \times k_W = 3 \times 3$.}
  \label{fig:pooling_example}
\end{figure}

Pooling operations reduce the spatial dimensions (width and height) of feature maps. This provides a degree of translation invariance and reduces the computational cost in subsequent layers. Pooling operates independently on each feature map. A pooling kernel of size $k_P \times k_P$ slides across the input map with a stride $\Sigma$. Common types are average pooling (AvgPool), which computes the mean of the values within the kernel window, and maximum pooling (MaxPool), which takes the maximum value.

In this paper, we apply MaxPool after the activation function in each convolutional layer ($\ell \in \{1, 2, 3\}$). The inputs to the pooling operation are the activated feature maps $\xi^\ell_j$. We use a pooling kernel size $k_P=2$ and a stride $\Sigma=2$. If an input map has matrix dimensions $N_{in} \times N_{in}$, the output map dimensions $N_{out} \times N_{out}$ are given by:
\begin{equation}
  N_{out} = \left\lfloor \frac{N_{in}-k_P}{\Sigma} \right\rfloor + 1.
\end{equation}
For $k_P=2, \Sigma=2$, this simplifies to $N_{out} = N_{in}/2$.

The output feature maps after pooling in layer $\ell$, denoted $\Xi^\ell_j$, are computed from the activated maps $\xi^\ell_j$. Using $s', t'$ for the output map, $s, t$ for the input map $\xi^\ell_j$), the maximum pooling operation is:
\begin{equation}
\Xi^\ell_j(s', t') = \max_{\substack{0 \leq \sigma < k_P \\ 0 \leq \tau < k_P}} \xi^\ell_j(\Sigma(s'-1) + 1 + \sigma,\ \Sigma(t'-1) + 1 + \tau), \quad \text{for } j=1, \dots, C_\ell.
\label{eq:maxpool_formula}
\end{equation}
With $k_P=2$ and $\Sigma=2$, this matches the formulae used in the main text, e.g., equation~(\ref{eq:layer1_Xi_pool}).

The average pooling equivalent (not used in the main text) would be:
\begin{equation}
  \Xi^\ell_j(s', t') = \frac{1}{k_P^2} \sum_{\sigma=0}^{k_P-1} \sum_{\tau=0}^{k_P-1} \xi^\ell_j(\Sigma(s'-1) + 1 + \sigma,\ \Sigma(t'-1) + 1 + \tau), \quad \text{for } j=1, \dots, C_\ell.
\end{equation}

\section{Notational analogy to fully connected neural networks}\label{sec:analogy_fcnn}

For conceptual comparison, we can draw an analogy between the operations in a convolutional layer (without pooling) and a fully connected neural network layer. Consider a simplified 1D case, where the output $y_j^\ell$ of neuron $j$ in layer $\ell$ of a fully connected neural network is typically computed as:
\begin{subequations}
\begin{align}
  y_j^\ell &= \mathcal{A}(z_j^\ell), \quad\text{where} \label{eq:fcnn_activation}\\
  z_j^\ell &= \sum_{k=1}^{n_{\ell-1}} w_{jk}^\ell y_k^{\ell-1} + b_j^\ell. \label{eq:fcnn_linear}
\end{align}
\end{subequations}
Here, $\mathcal{A}$ is the activation function, $z_j^\ell$ is the pre-activation value, $y_k^{\ell-1}$ are the activations from the previous layer ($\ell-1$), $w_{jk}^\ell$ is the weight connecting input neuron $k$ to output neuron $j$, $b_j^\ell$ is the bias for neuron $j$, and $n_{\ell-1}$ is the number of neurons in layer $\ell-1$.

In the convolution neural network context (Eqs. \ref{eq:layer2_zeta} and \ref{eq:layer2_xi}, ignoring pooling and 2D structure for simplicity), the computation for the $j$-th feature map involves:
\begin{subequations}
\begin{align}
  \xi^\ell_j &= \mathcal{A}(\zeta^\ell_j + b^\ell_j), \quad\text{where} \label{eq:cnn_analogy_activation}\\
  \zeta^\ell_j &= \sum_{k=1}^{C_{\ell-1}} (\Xi^{\ell-1}_k \star \mathsf{W}^\ell_{jk}). \label{eq:cnn_analogy_conv}
\end{align}
\end{subequations}
The key difference lies in equation~(\ref{eq:cnn_analogy_conv}) using a local cross-correlation ($\star$) operation with shared weights within the filter $\mathsf{W}^\ell_{jk}$, rather than a simple weighted sum (dot product) over the entire previous layer output as in equation~(\ref{eq:fcnn_linear}). However, both involve a form of weighted summation over inputs from the previous layer (summing over index $k$ up to the number of input features/neurons, $C_{\ell-1}$ or $n_{\ell-1}$), followed by a bias and activation function, highlighting the conceptual link.

\section{Visualisation of the second and third convolutional layers}\label{appendix:conv_layers}
The figures in this appendix show the complete feature extraction process for the same atomic density profile used in figure~\ref{fig:feature_progression}. Each set of three figures corresponds to one convolutional layer: pre-activation maps (cross-correlation outputs), post-activation maps (after ReLU and bias), and final feature maps (after maximum pooling).

\begin{figure*}[p!]
  
  \includegraphics[width=\linewidth]{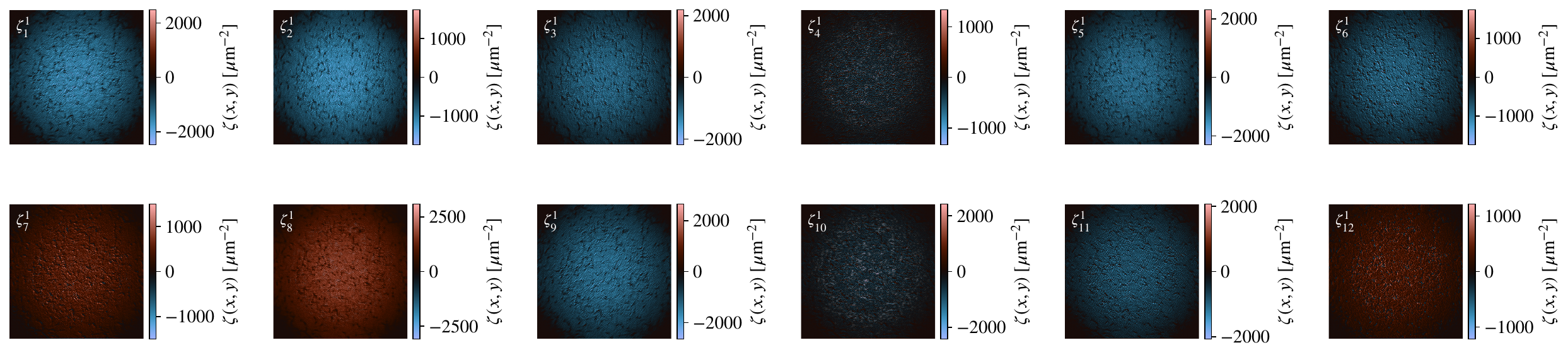}
  \caption{The cross-correlation, $\zeta^1$, of a {single image of an} atomic density, $\rho(x,y)$, with 12 weights kernels, $\mathsf{W}^1_i$, $i\in\{1,\cdots,12\}$, as determined by equation (\ref{eq:layer1_zeta}). The cross-correlation may result in positive or negative values, as indicated by the colour bars.}
  \label{fig:ch5:cross_correlation_1}
\end{figure*}

\begin{figure*}
  
  \includegraphics[width=\linewidth]{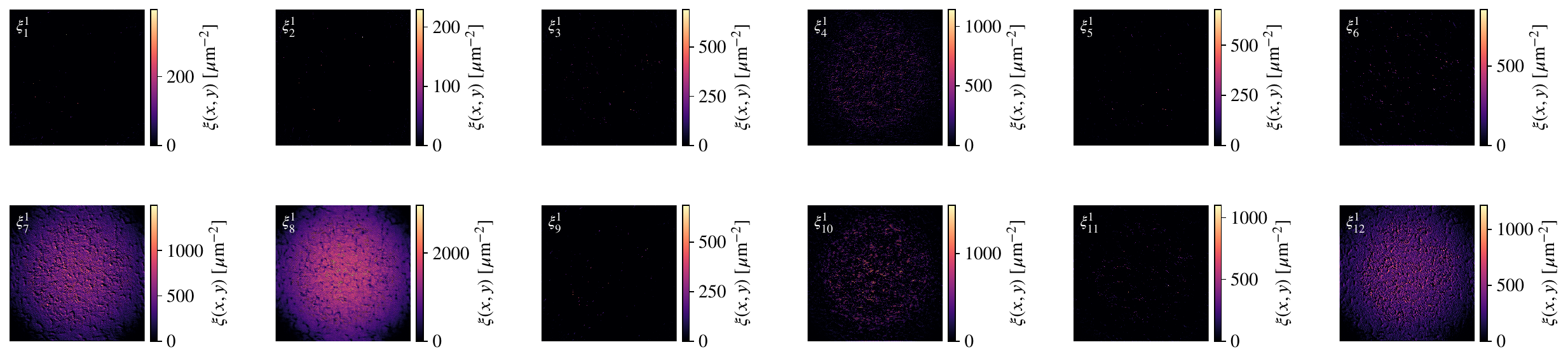}
  \caption[Chemical potential and temperature prediction model: first-layer feature maps]{$\xi^1$, the result of applying an activation function as determined by equation (\ref{eq:layer1_xi}).}
  \label{fig:ch5:conv_1}
\end{figure*}

\begin{figure*}
  
  \includegraphics[width=\linewidth]{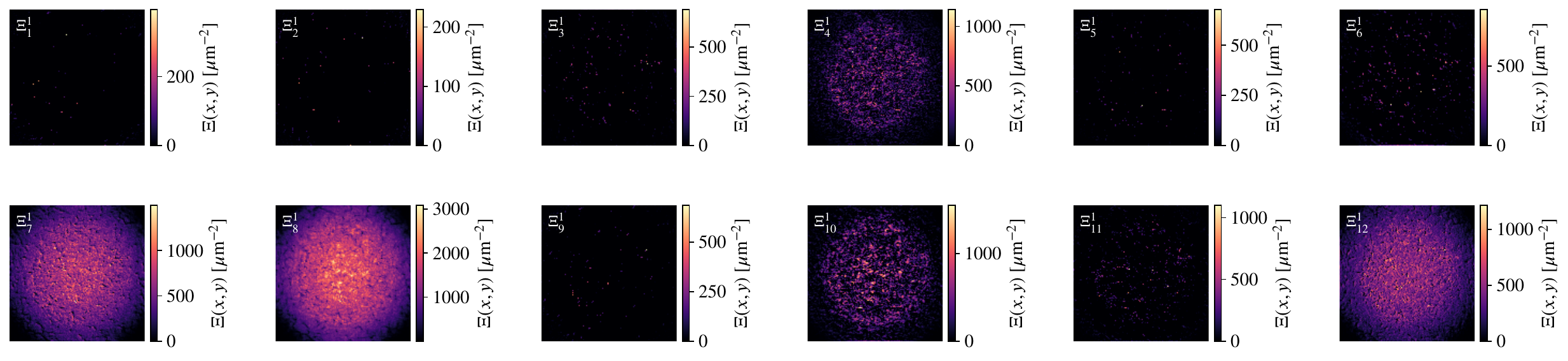}
  \caption[Chemical potential and temperature prediction model: first-layer maximally pooled features]{The maximally pooled features, and therefore the output of the first convolutional layer, $\Xi^1$, as determined by equation (\ref{eq:layer1_Xi_pool}).}
  \label{fig:ch5:pool_1}
\end{figure*}

\begin{figure*}
  
  \includegraphics[width=\linewidth]{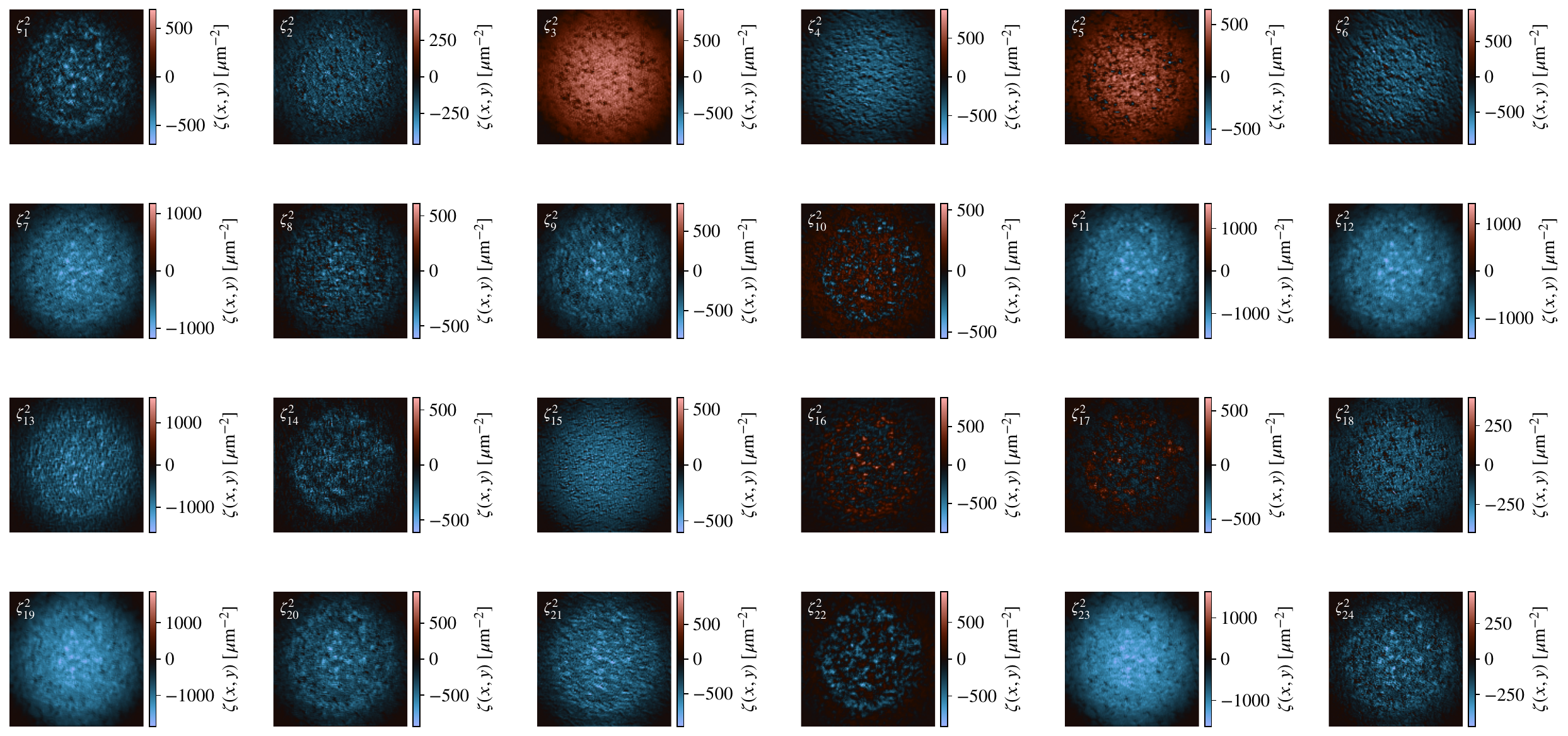}
  \caption[Chemical potential and temperature prediction model: second-layer cross-correlations]{The cross-correlation, $\zeta^2$, of the features extracted from the previous layer, $\Xi^1$, with 24 weights kernels, $\mathsf{W}^2_i$, $i\in\{1,\cdots,24\}$, as determined by equation (\ref{eq:layer2_zeta}). The cross-correlation may result in positive or negative values, as indicated by the colour bars.}
  \label{fig:ch5:cross_correlation_2}
\end{figure*}

\begin{figure*}
  
  \includegraphics[width=\linewidth]{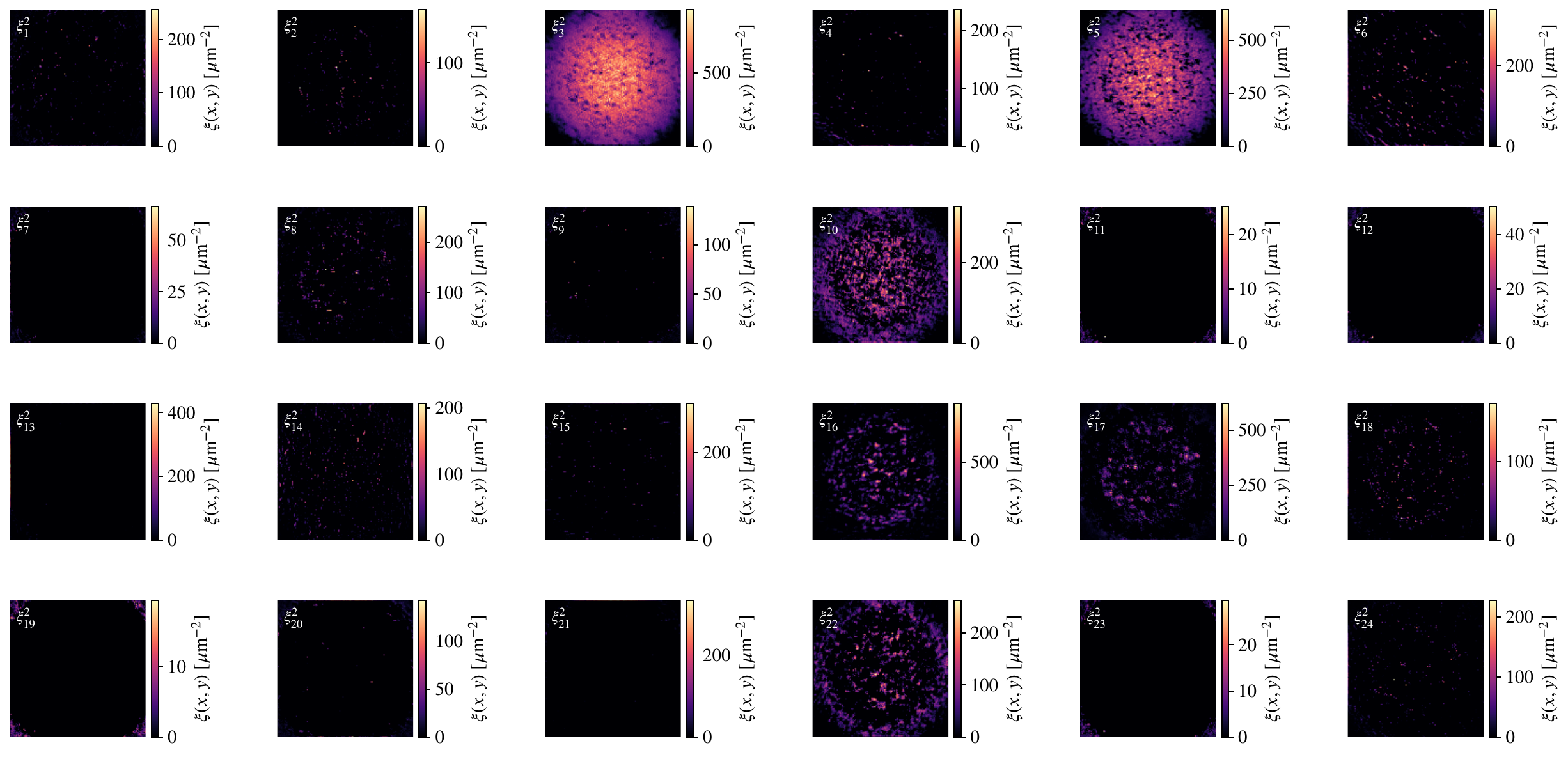}
  \caption[Chemical potential and temperature prediction model: second-layer feature maps]{The post-activations, $\xi^2$, as determined by equation (\ref{eq:layer2_xi}).}
  \label{fig:ch5:conv_2}
\end{figure*}

\begin{figure*}
  
  \includegraphics[width=\linewidth]{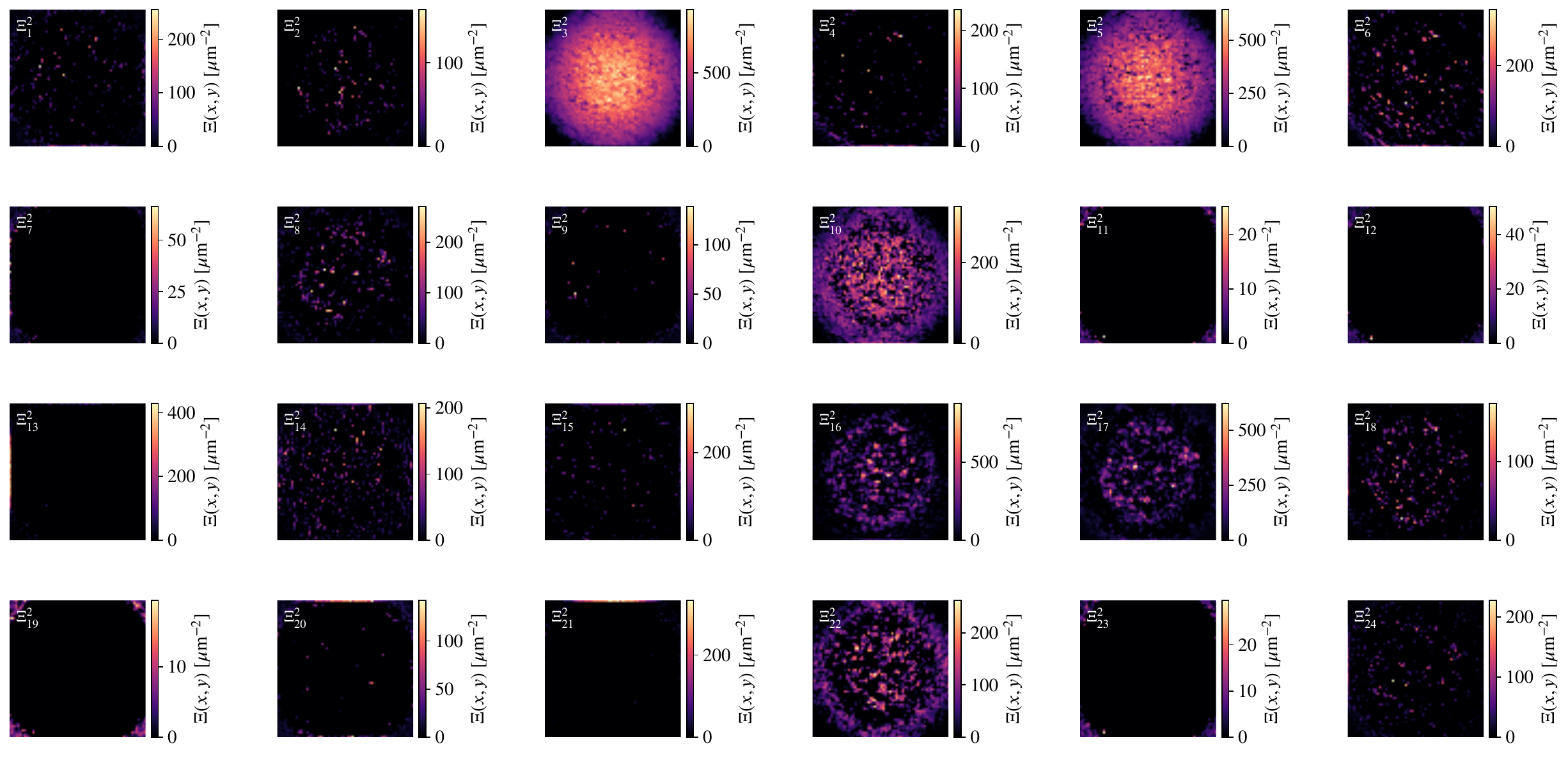}
  \caption[Chemical potential and temperature prediction model: second-layer maximally pooled features]{The maximally pooled features, and therefore the second layer outputs, $\Xi^2$, as determined by equation (\ref{eq:layer2_Xi_pool}).}
  \label{fig:ch5:pool_2}
\end{figure*}

\begin{figure*}
  
  \includegraphics[width=\linewidth]{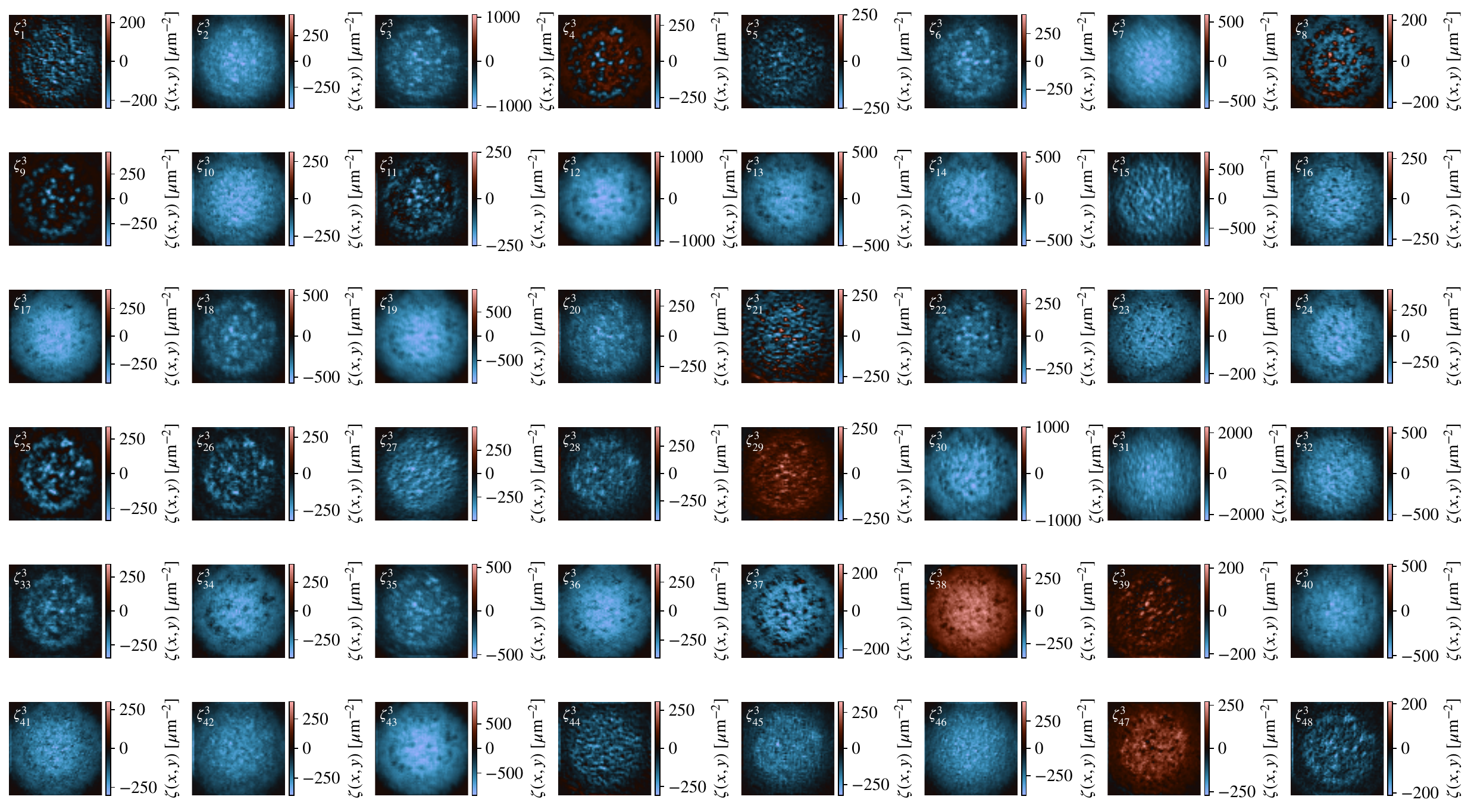}
  \caption[Chemical potential and temperature prediction model: third-layer cross-correlations]{The cross-correlation, $\zeta^3$, of the features extracted from the previous layer, $\Xi^2$, with 48 weights kernels, $\mathsf{W}^3_i$, $i\in\{1,\cdots,48\}$, as determined by equation (\ref{eq:layer3_zeta}). The cross-correlation may result in positive or negative values, as indicated by the colour bars.}
  \label{fig:ch5:cross_correlation_3}
\end{figure*}

\begin{figure*}
  
  \includegraphics[width=\linewidth]{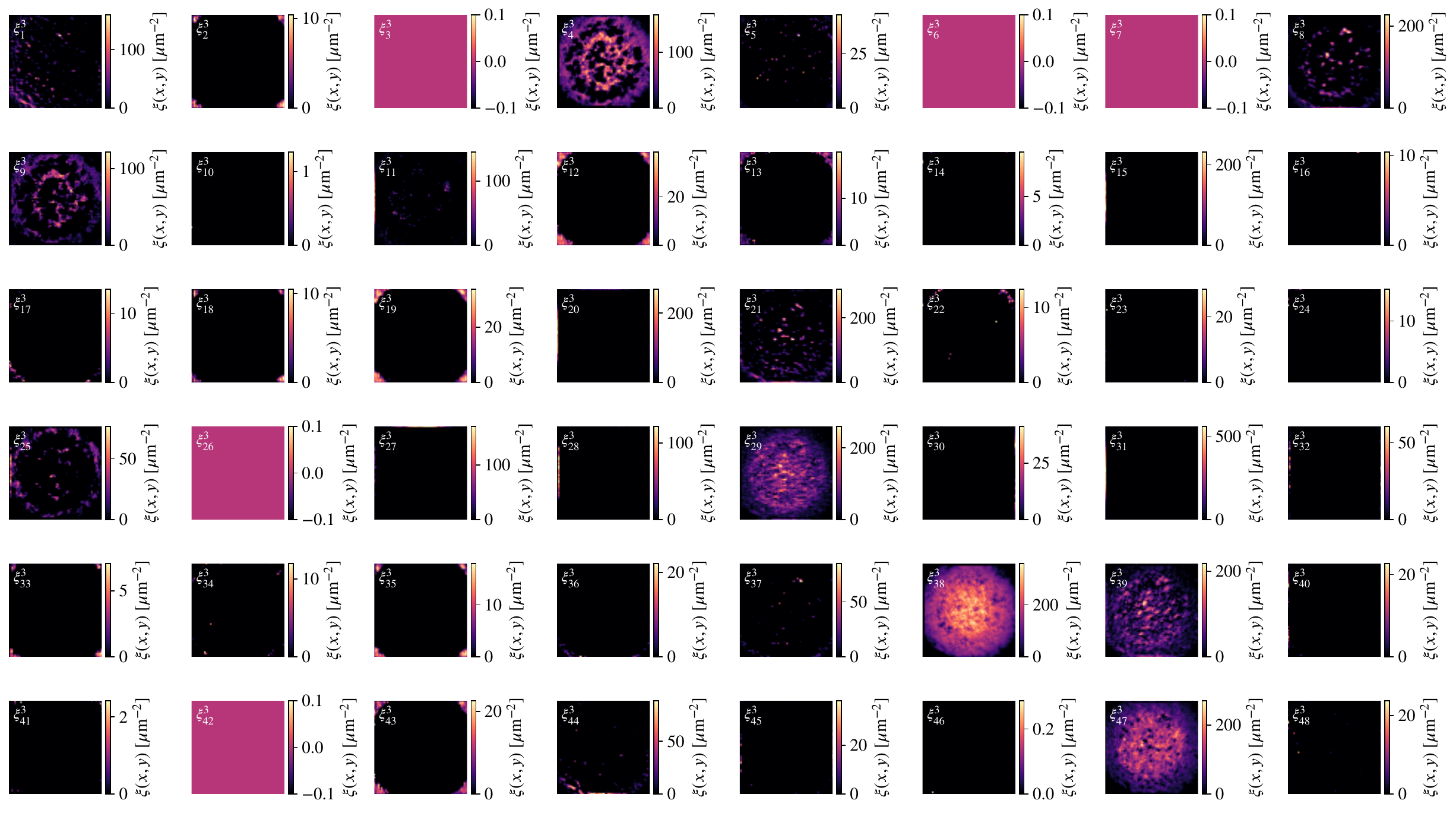}
  \caption[Chemical potential and temperature prediction model: third-layer feature maps]{The post-activations, $\xi^3$, as determined by equation (\ref{eq:layer3_xi}).}
  \label{fig:ch5:conv_3}
\end{figure*}

\begin{figure*}
  
  \includegraphics[width=\linewidth]{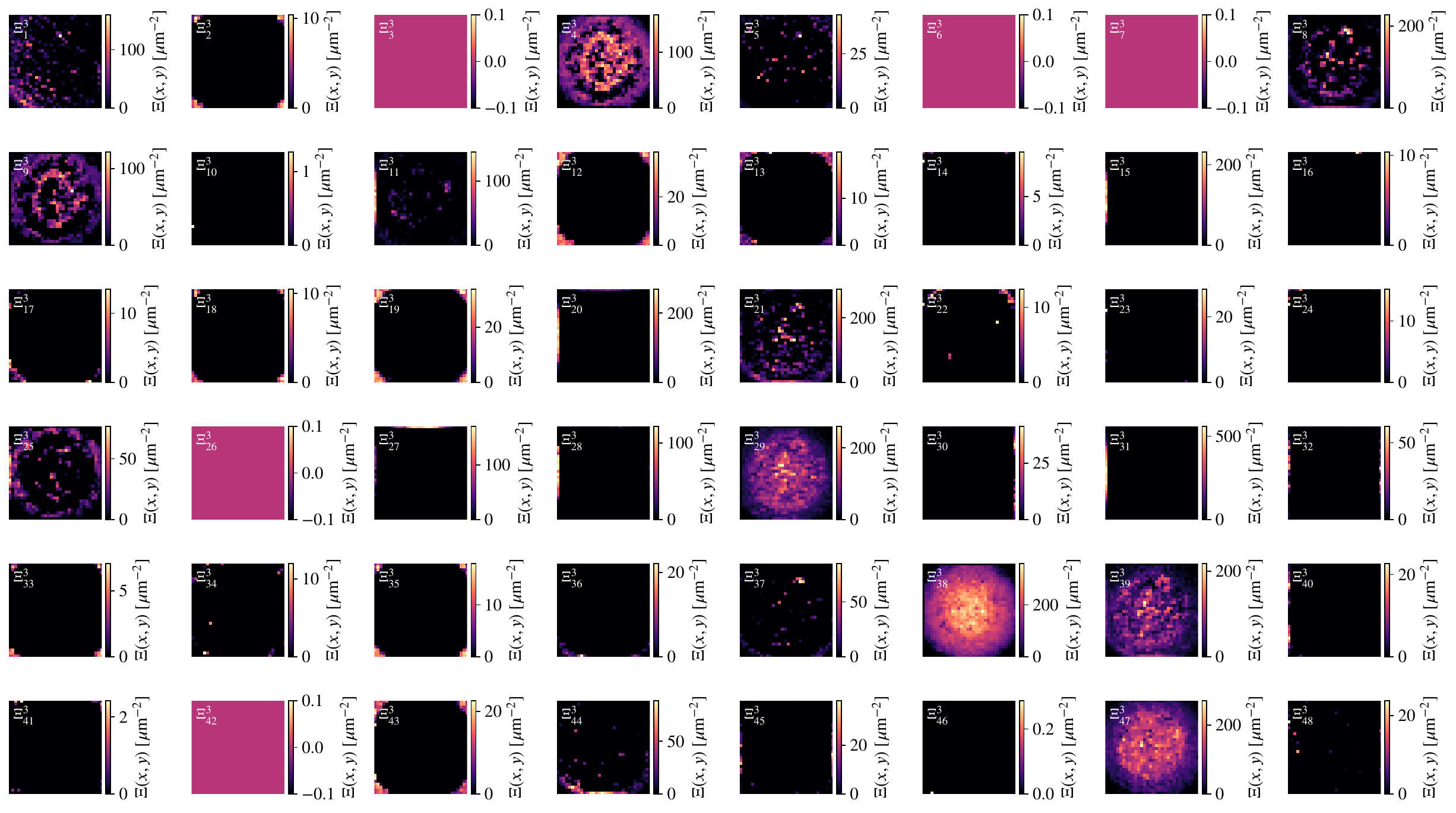}
  \caption[Chemical potential and temperature prediction model: third-layer maximally pooled features]{The maximally pooled features, and therefore the final output of the image processing pipeline, $\Xi^3$, as determined by equation (\ref{eq:layer3_Xi_pool}).}
  \label{fig:ch5:pool_3}
\end{figure*}

Figures \ref{fig:ch5:cross_correlation_1}--\ref{fig:ch5:pool_1} show the first layer outputs (12 features), which primarily capture high-density regions and large-scale structural information. Figures \ref{fig:ch5:cross_correlation_2}--\ref{fig:ch5:pool_2} show the second layer outputs (24 features), which learn intermediate-scale patterns. Figures \ref{fig:ch5:cross_correlation_3}--\ref{fig:ch5:pool_3} show the third layer outputs (48 features), which are sensitive to fine-scale fluctuations that encode temperature information.

\newpage

\bibliographystyle{iopart-num}
\bibliography{bib}

\end{document}